\documentclass[preprint, 11pt]{elsarticle} 

\usepackage{setspace}
\usepackage[margin=1in]{geometry} 

\usepackage{amsmath}
\usepackage{amsfonts}
\usepackage{amssymb}
\usepackage{amsthm}
\usepackage{bbm}
\usepackage{array}
\newcolumntype{C}[1]{>{\centering\let\newline\\\arraybackslash\hspace{0pt}}m{#1}}
\newcolumntype{L}[1]{>{\raggedleft\let\newline\\\arraybackslash\hspace{0pt}}m{#1}}
\newcolumntype{R}[1]{>{\raggedright\let\newline\\\arraybackslash\hspace{0pt}}m{#1}}

\usepackage{subcaption}
\usepackage{xcolor}
\usepackage{algorithm}
\usepackage{algpseudocode}
\usepackage{float}
\usepackage{hyperref}

\usepackage[para,online,flushleft]{threeparttable}

\theoremstyle{plain}
\newtheorem{theorem}{Theorem}[section]

\newtheorem{proposition}[theorem]{Proposition}

\theoremstyle{definition}
\newtheorem{definition}[theorem]{Definition}

\theoremstyle{remark}

\begin{document}

\begin{frontmatter}

\title{Healthcare Facility Assignment Using Real-Time Length-of-Stay Predictions: Queuing-Theoretic and Simulation-driven Machine Learning Approaches}

\author[inst1]{Najiya Fatma}

\affiliation[inst1]{Department of Mechanical Engineering, Indian Institute of Technology Delhi, New Delhi India}

\author[inst1]{Varun Ramamohan}


\begin{abstract}
Longer stays at healthcare facilities, driven by uncertain patient load, inefficient patient flow, and lack of real-time information about medical care, pose significant challenges for patients and healthcare providers. Providing patients with estimates of their expected real-time length of stay (RT-LOS), generated as a function of the operational state of the healthcare facility at their anticipated time of arrival (as opposed to estimates of average LOS), can help them make informed decisions regarding which facility to visit within a network. In this study, we develop a healthcare facility assignment (HFA) algorithm that assigns healthcare facilities to patients using RT-LOS predictions at facilities within the network of interest. We describe the generation of RT-LOS predictions via two methodologies: (a) an analytical queuing-theoretic approach, and (b) a hybrid simulation-driven machine learning approach. Because RT-LOS predictors are highly specific to the queuing system in question, we illustrate the development of RT-LOS predictors using both approaches by considering the outpatient experience at primary health centers. Via computational experiments, we compare outcomes from the implementation of the RT-HFA algorithm with both RT-LOS predictors to the case where patients visit the facility of their choice. Computational experiments also indicated that the RT-HFA algorithm substantially reduced patient wait times and LOS at congested facilities and led to more equitable utilization of medical resources at facilities across the network. Finally, we show numerically that the effectiveness of the RT-HFA algorithm in improving outcomes is contingent on the level of compliance with the assignment decision. 
\end{abstract}

\begin{keyword}
Length of stay prediction \sep real-time delay prediction \sep simulation-driven machine learning \sep healthcare facility assignment \sep queuing systems 
\end{keyword}

\end{frontmatter}

\section{Introduction}
\label{sec1}

Overcrowding is a growing concern at healthcare facilities worldwide, both in emergency departments and outpatient facilities, including those in India \citep{halder2025access}. Medical capacity expansion, flow optimization, congestion prediction, and patient diversion are common approaches to minimize delays and congestion at healthcare facilities \citep{duma2023real, chen2025optimal, shaposhnik2025dual, wang2025coordination}. Prior assignment of healthcare facilities to patients within a network has proven effective in balancing load distribution, reducing staff burden, and minimizing delays while enhancing patient experience \citep{arora2023probabilistic}. In emergency settings, a patient's real-time delay estimate - the immediate wait time before they receive care - serves as a key metric to initiate assignment within a network \citep{fatma2023patient}. In non-emergency or outpatient settings, patients typically interact with multiple subsystems (e.g., clinicians, nurses, pharmacists, etc.), making Length of Stay (LOS) - the total time spent in the system - a more relevant indicator of efficiency. For example, a person wishing to fit a visit to the doctor into the lunch break during their work day is likely to be much more interested in their expected LOS at the facility rather than only their expected wait time prior to seeing the doctor. Similarly, providing estimates of expected LOS can help healthcare administrators adjust staffing levels, reduce service time, or divert future arrivals to alleviate overcrowding \citep{gartner2020machine}. In this context, if real-time healthcare facility assignment decisions are to be performed using LOS estimates, real-time LOS (RT-LOS) estimates - i.e., the anticipated LOS of the patient at the facility as a function of the state of the facility at the anticipated time of arrival of the patient at the facility, as opposed to using average LOS estimates - represents the most up-to-date and `personalized' information that can be used for operational decision-making.




Estimating LOS on a real-time basis remains a challenging problem. Most empirical work on LOS prediction has primarily used historical data from the queuing system of interest (i.e., the facility) to train statistical or machine learning (ML) methods such as tree-based models and neural networks \citep{jain2024predicting,boff2025predicting}. Such models can struggle to adapt to dynamic healthcare environments characterized by unpredictable future arrivals and time-evolving patient conditions \citep{bravo2024interpretable,carmeli2023state}. Further, historical system state data (features) and the corresponding wait times (labels) for the queuing system under consideration are rarely available in sufficient amounts to train ML methods to achieve desired RT-LOS prediction accuracies \citep{baldwa2020combined}. On the other hand, from a queuing-theoretic standpoint, \cite{dong2019impact} showed that history-based queuing-theoretic real-time delay predictors create oscillations by providing time-lagged information. 

In this context, the objective of our study is as follows. We develop a framework via which real-time facility assignment decisions can be made via RT-LOS predictions. In generating the RT-LOS predictions required for implementation of the framework, we attempt to address the RT-LOS estimation issues described above with both the analytical queuing-theoretic and the data-driven statistical learning approaches. As part of this, we estimate the patient's LOS starting from the time at which the patient is expected to arrive at the facility of interest: i.e., at time $t + \delta$, where $t$ is the current time at the patient's point of origin and $\delta$ is the travel time to the facility. Given that the information available at $t$ only consists of system state information at $t$ (such as queue length and elapsed service time), we generated RT-LOS predictions at $t + \delta$ via two approaches. The first, an analytical queuing-theoretic (AQT) approach, extrapolates the system state information from $t$ to $t + \delta$ and then predicts, via queuing-theoretic delay prediction methods, the RT-LOS at $t+\delta$ as a function of the estimated expected system state at $t + \delta$. The extrapolation of system state information from $t$ to $t + \delta$ attempts to address the time lag issue. The second approach, which we call the Sim-ML method, addresses difficulties around data availability for queuing systems by utilizing a discrete-event simulation of the facility to generate system state (features, at $t$) and LOS (label, at $t + \delta$) data that we use to train ML-based RT-LOS predictors. We integrate these predictions into a real-time healthcare facility assignment (RT-HFA) algorithm that can inform patients regarding which facility they should visit in a network. 

Real-time delay - and therefore LOS - predictors are highly specific to the queueing system under consideration. Therefore, we illustrate the implementation of the AQT and SimML RT-LOS prediction methods and the RT-HFA method via a case study involving a network of primary health centers (PHCs) as found in the Indian public healthcare system. These are small clinics with one or two doctors and a few nurses, and provide primary care via outpatient consultations, and on a limited basis, serve patients requiring inpatient care and uncomplicated childbirth cases. The AQT RT-LOS predictor that we develop for this facility is predicated on a conceptualization of the queuing (sub)systems within the PHC as $M/G/n$, $G/G/n$, and multiclass queuing systems with non-preemptive priority for higher priority arrival classes. Therefore, our proposed AQT RT-LOS method, along with the queuing-theoretic system state extrapolation procedure, is broadly applicable to such queuing systems, and forms a key contribution of this study. 

The Sim-ML approach for RT-LOS prediction has also, in our knowledge, not previously been introduced in the literature. We also provide a straightforward analysis of the interaction between simulation and ML regressor errors for the Sim-ML approach. Further, in our knowledge, a comparison of this approach with AQT methods for either real-time delay or LOS prediction has not been done previously. Thus, these form key contributions of this study.

The paper is structured as follows. In Section~\ref{litrev}, we review the relevant literature, and in Section~\ref{rthfalgo}, we describe the RT-HFA algorithm and the PHCs for which the RT-LOS predictors are developed. In Section~\ref{lospred}, we present the development of the AQT and Sim-ML RT-LOS predictors, and in Section~\ref{rtlos_sec4}, we describe the computational experiments illustrating the implementation of the RT-HFA framework, and demonstrate the impact of patient compliance on the efficacy of the HFA algorithm. In Section~\ref{rtlos_sec5}, we summarize study findings, future research directions, and key managerial insights.

\section{Literature Review}
\label{litrev}

We first discuss studies that address healthcare facility assignment via the use of real-time information. We then review existing approaches to LOS prediction, with an emphasis on personalized, real-time estimation.

\subsection{Real-time healthcare facility assignment}

Real-time operational variable estimates, such as real-time delays, have long been utilized to enhance hospital operations and inform patient decision-making at healthcare facilities. However, most studies employing such methods fall short of showing how they can directly support healthcare facility assignment decisions. \cite{papi2016new} used LOS estimates to improve hospital bed planning. \cite{oakley2020symbiotic} used a symbiotic simulation/digital twin that includes conditional (on system state) LOS prediction to improve the hospital bed management performance. Though it does not use LOS estimates, the facility recommendation system NHSQuicker uses expected waiting times \citep{mustafee2018rh}.

\citet{bravo2024interpretable} utilized LOS predictions to determine whether the system is likely to become congested, that is, reach high occupancy levels, at a pre-specified point of time in the near future using queuing theory, simulation, and machine learning, and demonstrated the effectiveness of their proposed operational decision rules through a large-scale ICU model. \citet{westphal2022reducing} developed an application, MyED, that enabled individual patients to access their specific ED procedures and associated waiting times on their personal mobile phones in a medium-sized tertiary hospital. The authors incorporated queuing-theoretic features in machine learning methods to estimate individual patient waiting times for each specific ED procedure. 
 
\citet{zhang2020simulation} combined DES with an optimization model to improve patient assignments to clinical services across multiple units. 
\cite{song2015diseconomies} emphasized the value of effective queue management in reducing patient wait times and LOS, with simulation outputs showing high stability. \cite{dai2019inpatient} developed an efficient inpatient overflow management policy using a discrete-time, infinite-horizon Markov decision process that reduced unnecessary patient transfers by 20\% while maintaining similar congestion levels. \cite{li2016optimal} proposed physician assignments from hospitals to outreach clinics closer to patient residences, using a multi-commodity capacitated facility location problem, which was solved using column generation and local search heuristics.

An important consideration in healthcare facility assignment decisions involves the sharing of information regarding operational variables at each facility in the network~\citep{dong2019impact}. Effective communication among service-providing facilities helps alleviate congestion across the network, and the assignment decision must be made using information available from electronic management systems networked across all facilities~\citep{adjerid2018reducing, enayati2018real}. The real-time healthcare facility assignment problem proposed in the current work assumes the presence of a centralized information technology infrastructure that maintains information regarding operational statuses - such as elapsed service times in various queues within the facility - available to all facilities within the network. Examples of such IT systems can be found in the United Kingdom's National Health Service~\citep{mustafee2020providing}. \cite{fatma2023patient} also proposed and demonstrated computationally a centralized diversion mechanism for emergency patients across a multi-facility network that utilized real-time delay predictions at each facility in the network. However, as is realistic for emergency patients (e.g., transported in ambulances), patients in their study were diverted to the final destination facility after reaching or when they were already on their way to the original destination facility. 

Traditionally, metrics such as average wait time, bed occupancy, and queue lengths have been used to direct patients to alternative facilities within a healthcare network. To date, in our knowledge, no healthcare facility assignment method has explicitly used real-time length of stay (RT-LOS) estimates generated at the time a patient is expected to arrive at the facility. For non-emergency patients seeking consultation or diagnostic services on walk-in basis at a facility, LOS estimates at the facility, as opposed to wait time estimates at individual subsystems of the facility, are most relevant. Such systems are ubiquitous in low- and middle-income countries, especially in South Asia \citep{shoaib2022simulation}. Our study contributes to the healthcare facility assignment literature, and indeed more broadly to the real-time service facility assignment body of work. We demonstrate how RT-LOS estimates can inform facility assignment decisions for such service-seeking entities (patients) and (healthcare) facility networks. We now discuss the literature around LOS prediction and our contributions relative to the said body of work.

\subsection{Real-time length of stay prediction approaches}

Given the importance of length of stay prediction in healthcare OR/MS performance improvement studies, several studies focused on predicting LOS for emergency patients using statistical and machine learning techniques. \citet{papi2016new} employed a statistical Hypergamma density model to describe LOS distributions of patients across different departments in an Italian hospital over 5 years. \cite{grand2016regression} combined pseudo-observations with landmarking to construct generalized linear regression models for estimating expected LOS in multi-state models. \cite{zhang2019lognormal} used lognormal mixture distributions to model skewed, multimodal LOS data with outliers. There has been a surge in the use of data mining techniques for LOS prediction, especially in complex queuing systems where closed-form expressions are difficult to derive. \cite{boff2025predicting} predicted pediatric patients' LOS in hospitals using ML-based algorithms.


Data-driven methods for real-time LOS estimation typically involve training ML methods directly on queuing system and patient characteristic datasets. However, sufficiently rich datasets that can be leveraged to yield acceptable prediction accuracies can be challenging to obtain, and in this context, we present a methodology that uses a validated discrete-event simulation of the queuing system in question to generate the system state data to train ML methods in question. This approach has previously been used for real-time delay prediction in the healthcare operations context \citep{baldwa2020combined,fatma2024simulation} and also on a limited basis for estimating congestion risk in intensive care units \citep{bravo2024interpretable}. In our study, we use a validated DES to record system state data (features) at the current time ($t$, when the patient is making their decision regarding which facility to visit from their point of origin) and their actual LOS experienced starting from their time of arrival ($t + \delta$) at the facility (the prediction label). ML regressors - the RT-LOS predictors - are trained on this dataset generated from the simulation. To our knowledge, there is no other study that has generated data-driven RT-LOS predictions in this realistic context, where these predictions are generated at the time when the facility choice is made.

As discussed briefly in Section~\ref{sec1}, from a LOS prediction standpoint, the research contributions of this study advance queuing-theoretic as well as data-driven approaches for RT-LOS estimation. The RT-LOS prediction approach uses real-time delay predictions. Real-time delay prediction has been an active area of research \citep{ibrahim2018managing,ibrahim2017does,fatma2024simulation,baldwa2020combined}, with a wide variety of queuing-theoretic and data-driven approaches. However, in our knowledge, there is no study that attempts to estimate delay, and subsequently LOS, on a real-time basis at a future time point by extrapolating system state information recorded at the current time to the future time point. Our study, by proposing and demonstrating a procedure for this purpose for $M/G/n$ and $G/G/n$ systems, as well as multi-class systems with non-preemptive priority, contributes to both queuing-theoretic real-time delay and LOS prediction.

While \citet{oakley2020symbiotic} used a symbiotic simulation to generate conditional LOS distributions based on real-time system states, our work introduces: (a) an analytical queuing-theoretic (AQT) predictor that uses real-time delay predictions, and (b) a simulation-driven machine learning (Sim-ML) approach for real-time LOS predictions. Our proposed AQT and Sim-ML approaches for LOS estimation are likely to be easier to deploy and faster in generating LOS predictions than real-time simulations. This is because they require only a single function evaluation for generating real-time LOS prediction, whereas the real-time simulation may need to be executed multiple times to generate the (average) real-time LOS predictions. We compare the effectiveness of the AQT and Sim-ML approaches on network operational outcomes, and also provide an analytical treatment of how the simulation and ML regression errors interact, both of which have previously not been done.

\section{Real-time Healthcare Facility Assignment: Algorithm and Case Study Overview}
\label{rthfalgo}

\subsection{The Algorithm}
\label{sec:rthfalg}

We present the real-time healthcare facility assignment mechanism in Algorithm~\ref{alg:the_alg}, referred to henceforth as RT-HFA, that uses real-time LOS predictions at a subset of facilities under consideration to drive the assignment decision. 

\begin{algorithm}[htbp]
	\caption{RT-HFA: Real-time healthcare facility assignment algorithm.}
	\label{alg:the_alg}
	\begin{algorithmic}[1]
        \State Initialize with empty LOS list $\mathcal{L} = [~]$
		\State Patient is at point of origin at time $t$
        \State Select $m~(m \leq k)$ out of $k$ facilities for facility assignment
        \Comment{/*Example: $m$ out of $k$ facilities nearest to patient*/}
        \For{$j = 1 \text{ to } m$}
		\State Record system state information ($s_{j(t)}$) at time $t$ for $j^{th}$ facility
		\State Estimate expected travel time ($\delta_{j}$) of patient to $j^{th}$ facility
		\State Estimate expected future system state $s_{j(t+\delta_j)}$ as $f(s_{j(t)})$
		\State Generate, at time $t$, real-time LOS prediction $L_{j(t+\delta_j)}$ as $g(s_{j(t+\delta)})$
		\State Store the quantity $\delta_j + L_{j(t+\delta_j)}$ in LOS list $\mathcal{L}$
        \EndFor
		\State Compute $j^* = \arg \underset{L_{j(t+\delta_j)}}{\min}~\mathcal{L}$ 
        \State Facility $j^*$ is assigned to patient
	\end{algorithmic}
\end{algorithm}


In the RT-HFA mechanism presented in Algorithm~\ref{alg:the_alg}, assuming $k ~ (\geq 2)$ facilities are available for assignment, $m~(\leq k)$ facilities are considered for assignment to the patient under consideration. These can be selected as, for example, the $m$ nearest facilities to the patient. We assume that, at current time $t$, the patient accesses an information technology system - either a web-based or smartphone application - that performs the facility assignment based on Algorithm~\ref{alg:the_alg}. It is assumed that the operational state of each facility under consideration - at time $t$ - is linked to this application, and that it is provided as input to Algorithm~\ref{alg:the_alg}. Such an application can be linked to the enterprise resource planning (ERP) software used by the facility - for example, the ERP system that checks patients into the facility can record arrival of patients into the facility and the system within the doctor's or nurse's stations can be used to record consultation start and end times. 

The RT-HFA mechanism in Algorithm~\ref{alg:the_alg} performs the assignment at time $t$ based on the operational states $s_{j(t)}~(j = 1 \text{ to } m)$ of the $m$ facilities at time $t$; however, it does so by estimating the expected real-time LOS $L_{j(t + \delta_j)}$ at time $t+\delta_j$. This latter computation is done by estimating the future operational states $s_{j(t + \delta_j)}$ of the $m$ facilities as functions of the $s_{j(t)}$. The RT-HFA algorithm then computes $\delta_j + L_{j(t+\delta_j)}$ for each facility so that the distance to the facility as well as its estimated real-time LOS are considered while making the assignment decision. Section \ref{lospred} describes approaches to perform the computations in lines 7 and 8 in Algorithm~\ref{alg:the_alg}. 




Two aspects of the RT-HFA mechanism merit elaboration. First, in our conception, it is likely that the RT-HFA mechanism will be used to provide the most `convenient' option based on operational considerations to a patient; however, the final choice of which facility to visit will likely still remain with the patient. Therefore, in evaluating the operational impact of the RT-HFA mechanism, patient compliance rates with the facility recommendation must also be considered. This is demonstrated via computational experiments in Section~\ref{rtlos_sec4}.

Second, the RT-HFA algorithm takes a centralized approach, meaning the operational states of the system of all facilities involved in the network are considered in performing the assignment. An optimal assignment choice made independently by taking into consideration the operational details of only one healthcare facility, without considering others in the network, might lead to congestion elsewhere, impacting the overall network performance \citep{deo2011centralized}.


To generate LOS estimates at times $t+\delta$, we employ and compare two techniques: the AQT and Sim-ML methods. The AQT and the Sim-ML approaches will typically have to be developed specifically for the queuing system under consideration. In this study, we illustrate the development of RT-LOS predictors under these approaches for the queuing systems at healthcare facilities relevant to a specific case study: RT-HFA for a network of primary health centers (PHCs). We now discuss the relevant operations at PHCs, and also show how, with reference to a real-world healthcare operations management situation, developing an HFA mechanism based on RT-LOS predictions can become relevant. Subsequent to this, we discuss the AQT approach in Section~\ref{secsimaqt} and the Sim-ML approach in Section~\ref{secsimml}.

\subsection{Case Study Overview: Primary Health Centers}
\label{background}



PHCs form the initial point of medical contact for the public with trained doctors, and provide care to outpatients, inpatients on a limited basis, childbirth patients (uncomplicated cases), and pregnant women seeking antenatal care. The outpatient department (OPD) operates for eight hours a day, while inpatient and childbirth services are accessible on a 24×7 basis. The Indian government has recently undertaken an initiative to expand and upgrade existing PHCs \citep{blanchard2021vision}, which will likely yield an increase in patient load at these PHCs given the established link between the increased quality of care, improved infrastructure, and higher demand \citep{rao2018quality}. 

A typical PHC has one or two general physicians, a duty nurse caring for inpatients and childbirth patients, and a nurse who assists physicians with outpatient care, with a focus on noncommunicable diseases (henceforth referred to as the `NCD' nurse). The facility has four to six inpatient beds and a single-bed labor room specifically for admitting childbirth patients. Additionally, PHCs have an in-house clinical laboratory for conducting routine laboratory investigations, and a pharmacy for dispensing free medications and handling patient registrations at the facility. We present an overview of the patent flow within the healthcare facility in Figure~\ref{Fig1a} below.

\begin{figure}[htb]
    \centering
    \includegraphics[width=0.90 \textwidth]{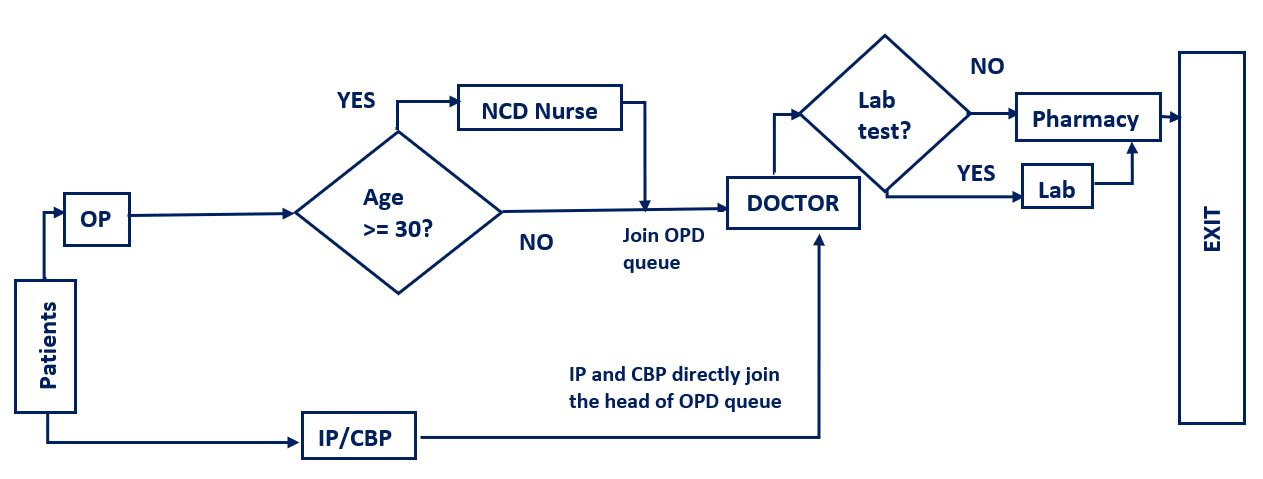}
    \caption{Patient flow at PHC. OP: Outpatients, IP: Inpatients, CBP: Childbirth patients.}
    \label{Fig1a}
\end{figure}

Upon arrival, outpatients below 30 years directly join the queue for consulting with the doctor. Patients aged 30 and above first consult the NCD nurse for hypertension and high blood glucose level screening as part of a preventive program for lifestyle-related diseases before seeing the doctor. After consulting with the doctor, about fifty percent of outpatients are directed to the in-house laboratory for completing their prescribed investigations. All the outpatients exit via the pharmacy to register their visit and obtain medication if required.

Inpatients arriving during outpatient hours directly consult the doctor before being admitted to the inpatient ward. Outside outpatient hours, inpatients are attended to by the duty nurse on call. Childbirth patients are initially seen by the doctors during outpatient hours, who then request a bed for them in the labor room. Outside OPD hours, childbirth patients are attended to by the duty nurse while the rest of their patient flow remains the same. Both inpatients and childbirth patients have non-preemptive priority over outpatients during outpatient hours, implying these patients move to the head of the doctor's queue upon their arrival (ahead of outpatients already in the queue). Table~\ref{tab1} reports details of the service processes at these medical resources. All arrival processes - for outpatients, inpatients, and childbirth patients - were modelled as stationary Poisson processes. The process of parameterizing these arrival and service processes at the PHC is described in detail in \cite{shoaib2022simulation}.


\begin{table}[htbp]
  \centering
  \caption{Input parameter estimates at PHC.}
    \begin{tabular}{|p{11.855em}|l|}
    \hline
    Input parameters & \multicolumn{1}{p{10.855em}|}{Arrival / service process model (in minutes)} \\
    \hline
    Interarrival times  &  \\
    Outpatients &  \\
    a. PHC$_{1}$ & \multicolumn{1}{p{10.855em}|}{$Exp(9)$} \\
    b. PHC$_{2}$ & \multicolumn{1}{p{10.855em}|}{$Exp(2)$ } \\
    Inpatients & \multicolumn{1}{p{10.855em}|}{$Exp(2880)$ }   \\
    Childbirth patients & \multicolumn{1}{p{10.855em}|}{$Exp(2880)$ }   \\
    \hline
    Service times at the doctor  &  \\
    a. Outpatient & \multicolumn{1}{p{10.855em}|}{$N(0.87, 0.21^{2})$} \\
    b. Inpatient & \multicolumn{1}{p{10.855em}|}{$U(10,30)$} \\
    c. Childbirth patient & \multicolumn{1}{p{10.855em}|}{$U(30,60)$} \\
    \hline
    NCD nurse service time & $U(2, 5)$ \\
    \hline
    Laboratory service time & $N(3.45, 0.83^{2})$ \\
    \hline
    Pharmacy service time & $N(2.08, 0.72^{2})$ \\
    \hline
    \end{tabular}%
  \label{tab1}%
\end{table}%

In this work, we primarily focus on the flow of outpatients within PHCs. The flow of childbirth patients and inpatients has been previously addressed in \cite{fatma2023patient}, which discussed diverting these emergency patients to other (destination) facilities after they reach their original destination using real-time delay predictions. However, for outpatients, and indeed for non-emergency (walk-in) patients, we see that healthcare facility assignment, rather than diversion, is more meaningful. Diversion based on wait time to receive care is appropriate for emergency patients, and hence has been used in such scenarios. However, for facilities in a network that serve walk-in non-emergency patients (as is common in public health facilities across India), patients may wish to plan their schedules and healthcare-seeking visits based not only on their wait times to consult with the doctor, but also on the total amount of time they may spend at the facility. 


Using the patient flow outlined in Figure~\ref{Fig1a} and the parameter estimates in Table~\ref{tab1}, we constructed a discrete event simulation (DES) model of the patient flow operations at two PHCs. We programmed the simulation in the Python programming language, using the \textit{Salabim} package within the IntelliJ IDEA integrated environment. The simulation experiments were performed on a workstation equipped with an Intel \textit{i}7 processor and a 64-bit Microsoft Windows operating system with 16 gigabytes of memory. We simulated 365 days of PHC operations, with a 180 day warm-up period, and collected data from 50 replications, with key operational outcomes for both PHCs reported in Table~\ref{tab2}.

\begin{table}[htbp]
	\begin{threeparttable}
	\centering
	\caption{Operational outcomes across 2 PHCs.}
	\begin{tabular}{|p{6.07em}|p{7.07em}|p{7.5em}|p{10.855em}|}
		\hline
		{Outcomes} & PHC$_1^{+}$ & PHC$_2^{+}$  & {Percentage difference} \\
		   & 9/1/1/1/1/1/1 & 2/1/1/1/1/1/1 & {($\Delta_{net(o)}$)} \\
		\hline
		$\rho_{doc}$ & 0.463 (0.003) & 1.085 (0.004) & 57.31\% (0.38) \\
		\hline
		$\rho_{NCD}$ & 0.513 (0.005) & 1.472 (0.017) & 65.11\% (0.45) \\
		\hline
		$\rho_{phar}$ & 0.383 (0.003) & 1.704 (0.006) & 77.52\% (0.22) \\
		\hline
		$\rho_{lab}$ & 0.317 (0.003) & 1.414 (0.006) & 77.56\% (0.25) \\
		\hline
		$w_{opd}$ & 0.367 (0.081) & 2.843 (0.185) & 51.68\% (4.61) \\
		\hline
		$w_{phar}$ & 0.353 (0.013) & 35.82 (0.729) & 99.01\% (0.04) \\
		\hline
		$w_{lab}$ & 0.522 (0.010) & 18.58 (0.599) & 97.18\% (0.00) \\
		\hline
		$w_{NCD}$ & 0.434 (0.018) & 9.417 (0.387) & 95.37\% (0.27) \\
		\hline
		$LOS$ & 8.582 (0.123) & 58.492 (1.20) & 85.32\% (0.39) \\
		\hline
	\end{tabular}%
			\footnotesize{$^{+}$Outpatient interarrival time (in minutes)/number of doctors/number of NCD nurses/
			number of laboratory technicians/number of pharmacists. $^{*}$ $\rho$$_{s}$ are dimensionless fractions, $w$, and $LOS$ are in minutes. $^{*}$$\rho_{doc}$: doctor's utilization, $\rho_{NCD}$: NCD nurse's utilization, $\rho_{phar}$: pharmacist's utilization, $\rho_{lab}$: laboratory technician's utilization, $w_{opd}$: OPD queue wait time, $w_{phar}$: pharmacy queue wait time, $w_{lab}$: laboratory queue wait time, $w_{NCD}$: NCD nurse queue wait time, $\Delta_{net}$: difference between individual facility-specific operational outcomes $o$. }
		\label{tab2}%
		\end{threeparttable}
\end{table}

Under conditions of high outpatient demand at PHC$_{2}$ relative to PHC$_{1}$, we see a significant disparity in operational outcomes across both facilities. Given the short consultation times with the doctor and the NCD nurse at the PHC relative to outpatient interarrival times, the wait times at these subsystems are low at both PHCs. Table~\ref{tab2} shows that the majority of the resources in PHC$_{2}$ have higher levels of utilization, resulting in longer wait times for most patients. The wait times for outpatients in each queuing subsystem within PHC$_{2}$ are notably longer. For example, patients in PHC$_{2}$ typically wait nearly 20 minutes at the laboratory, compared to less than a minute at PHC$_1$. Every outpatient visiting the pharmacy here must wait for over thirty minutes on average before leaving the healthcare facility. All of these contribute substantially to higher overall outpatient LOSs at PHC$_{2}$ relative to PHC$_{1}$: that is, nearly an hour at PHC$_{2}$, compared to less than 10 minutes at PHC$_1$. These observations illustrate why, for such scenarios, an HFA mechanism based on RT-LOS predictions is likely to yield more meaningful improvements in operational outcomes than if it were based on real-time delay or wait time predictions. 


\section{Real-time Length of Stay Prediction}
\label{lospred}

\subsection{Analytical Queuing-theoretic Approach}
\label{secsimaqt}

We now describe the generation of real-time LOS predictions for outpatients at the PHC using the AQT approach. The AQT real-time LOS prediction model that we present in this section is based on the nature of the queuing systems suitable for modeling outpatient flow at PHCs. The NCD nurse, laboratory, and pharmacy service stations at PHCs can be approximated as $M/G/1$ systems. The care process at the doctor for outpatients is different, since doctors attend to childbirth patients as well as inpatients, both of which have non-preemptive priority over outpatients, meaning that they just move to the head of the queue, and do not interrupt the service of a patient receiving care. Therefore, the AQT real-time LOS predictor that we present can be applied directly to generate real-time LOS predictions for $M/G/1$ systems, and with minimal modification to general $M/G/n$ and $G/G/n$ systems. That said, the overarching AQT real-time LOS prediction approach - in particular, its development based on real-time delay prediction, and the approach for real-time LOS prediction at future time points based on current state information - is likely to be applicable for general queuing systems.  

The AQT prediction model utilizes system-state information such as queue length and elapsed service time at each subsystem within the PHC (e.g., the NCD nurse station, the doctor's station - see Figure~\ref{Fig1a}) to generate LOS predictions. The real-time LOS at the PHC will be a linear combination of the real-time LOS estimates at each of these subsystems. This is expressed in equation \ref{eq:lostot} below. 

\begin{equation}
	\label{eq:lostot}
    \begin{aligned}
	&LOS = w^T LOS_{phc}\\
    & \text {For a patient who visits all four PHC subsystems, we have: } w = \mathbbm{1} \in \mathbb{R}^4
    \end{aligned}
\end{equation}

In equation~\ref{eq:lostot}, $LOS_{phc} \in \mathbb{R}^4$ is a vector containing the real-time LOS estimates for the four subsystems of relevance to an outpatient, and $w$ is the vector of weights ($w_i \in [0,1] \text{ and } i \in I = \{n, d, l, p\}$) for the linear combination in equation~\ref{eq:lostot}. $I$ represents the set of subsystems relevant to an outpatient's possible trajectories through the PHC, where $n, d, l \text { and } p$ represent the NCD nurse, doctor, laboratory, and pharmacy subsystems, respectively. We recall here that the LOS estimation for an outpatient is made at $t +\delta$, the point at which they arrive at the PHC, and before their consultation with the doctor takes place. This means that information regarding whether they will visit the laboratory is not available at the time of LOS estimation, and hence $w_l = p_l$, the probability of visiting the laboratory for all patients (equal to 0.5 for the purposes of our study). However, given that visiting the NCD nurse is mandatory for all patients aged less than 30 years, and it is assumed that patient age information is available to the LOS prediction system, $w_n$ can be set to 0 or 1 based on the patient age. All patients visit the doctor and the pharmacy, and hence $w_d = w_p = 1$ for all patients. Thus, the vector of weights $w$ can be determined for each patient based on their characteristics. Algorithm~\ref{alg:loscalc} depicts this process.

\begin{algorithm}[htb]
	\caption{Calculation of the total LOS at a PHC.}
	\label{alg:loscalc}
	\begin{algorithmic}                      
		\State Initialize with the ordered list $I = [n, d, l, p]$, empty lists $w = [~], LOS_{phc} = [~]$. 
		\For{$i \in I$}
                \If{Patient Age $\geq$ 30 years AND $i = n$:}
                \State Estimate $LOS_i$
                \State Append $LOS_i$ to $LOS_{phc}$
                \State Append $w_i = 1$ to $w$
                \ElsIf{$i = n$:}
                \State $LOS_i = 0$
                \State Append $w_i = 0$ to $w$
                \State Append $LOS_i$ to $LOS_{phc}$
                \Else
                \State Estimate $LOS_i$
                \State Append $LOS_i$ to $LOS_{phc}$
                \If{$i = l$:}
                \State Append $w_i = 0.5$ to $w$
                \Else 
                \State Append $w_i = 1.0$ to $w$ 
                \EndIf
                \EndIf 
		\EndFor
        \State $LOS = w^T LOS_{phc}$
	\end{algorithmic}
\end{algorithm}

We now describe the general procedure for generating real-time LOS predictions relevant to the queuing system of interest in this study. 

\subsubsection{General AQT Real-time LOS Prediction Approach for Healthcare Facility Assignment}
\label{sec:aqtgen}

Each of the RT-LOS estimates $LOS_i$ in Algorithm \ref{alg:loscalc} is estimated via the AQT approach as the sum of the real-time delay prediction for the patient at the subsystem in question and the average service time at the system. Thus, it is the real-time delay predictions that drive the estimation of the real-time LOS. This is expressed in equation~\ref{eq:lossub} below. 

\begin{equation}
	\label{eq:lossub}
	LOS_{i,t} = d_{i,t} + E[X], \forall~i \in I 
\end{equation}

In equation~\ref{eq:lossub}, we add the $t$ suffix to the expected real-time LOS to indicate that the estimation is done on a real-time basis as a function of the state of the system $i~(i\in I)$ at time $t$. $d_{i,t}$ is the expected real-time delay for the patient at the subsystem $i \in I$ and at time $t$, and $E[X]$ is the expected service time. The expected real-time delay $d_{i,t}$, expressed below in \ref{eq:wtpred}, is a function of the number of patients ahead in the queue when the patient under consideration enters the queue ($L_{q(i,t)}$) and the expected remaining service time $w(x_t)$ for the patient currently receiving service at the queuing system, where $x_t$ is the elapsed service time of the entity receiving service at time $t$. 

\begin{equation}
	\label{eq:wtpred}
	d_{i,t} = L_{q(i,t)} E[X] + w(x_t)
\end{equation}


For a single-server queuing system with Poisson arrivals, $L_{q(i,t)}$ and $x_t$ together describe the operational \textit{state} of system $i$ at time $t$, which we can denote as the tuple $s_{i,t} = (L_{q(i,t)}, x_t)$. 


From equation~\ref{eq:wtpred}, $d_{i,t}$ is a function of $w(x_t)$ and $L_{q(i,t)}$. $L_{q(i,t)}$ is typically available as part of system state information. However, as will be discussed shortly, for our specific case, the goal is to estimate the length of the queue at the future point in time $t+\delta$ ($L_{q(i, t+\delta)}$), when the patient is expected to arrive at the facility, as a function of $L_{q(i, t)}$. The expected remaining service time $w(x_t)$ must be estimated as a function of the other system state descriptor $x_t$. 

For a queuing system with generally distributed service times, the expected remaining service time of a server $w(x_t)$ is traditionally estimated using equation~\ref{eq:wgen} below. In equation~\ref{eq:wgen}, $V$ is the random variable representing the remaining service time of the patient currently in service given the elapsed service time $x$, and $v$ is a realization of $V$. $X$ is the random variable representing the service time, and $G$ is the $cdf$ of the service time.

\begin{equation}
	\label{eq:wgen}
	\begin{aligned}
	&P(V \leq v | x) = \frac{P(x \leq X \leq v + x)}{P(X \geq x)}\\
	& \implies G(v|x) =  \frac{G(v + x) - G(x)}{1-G(x)}
\end{aligned}
\end{equation}

The expected value of the remaining service time $w(x)$ is then estimated by calculating the $pdf$ $g(v|x)$ from equation~\ref{eq:wgen}. However, this calculation can become non-trivial depending upon the nature of the distribution $G(v|x)$. For example, determining $w(x)$ for a triangular distribution involves working with a piecewise continuous $cdf$, and for the Gaussian distribution requires numerical integration involving the error function \citep{fatma2023patient}. Therefore, we used an alternate approximate remaining service time predictor developed for such systems with symmetric and unimodal service time distributions. This remaining service time predictor is considerably easier to compute and implement, and can also be extended to multi-class multi-server queuing systems. This predictor, proposed in \cite{fatma2023patient}, is given below. 

\begin{equation}
	\label{eqrstgen}
	w(x) =
	\begin{cases}
		G^{-1}{(0.5)}-x ,        & \text{$ 0\le x < G^{-1}{(0.5)}$} \\
		G^{-1}{(0.75)}-x,       & \text{$ G^{-1}{(0.5)}\le x< G^{-1}{(0.75)}$} \\
		\frac{G^{-1}(ext) - x}{2},          & G^{-1}(0.75) \le x\le G^{-1}(ext) \\
		
	\end{cases}
\end{equation}

Here, the estimation of $w(x)$ depends on where the patient's elapsed service time $x$ falls within the service time distribution, with respect to $G^{-1}{(0.5)}, G^{-1}{(0.75)}$ (the 50$^{th}$ and 75$^{th}$ quantiles) and an extreme quantile $G^{-1}{(ext)}$ (e.g., with $ext = 0.99$) or $G^{-1}(1)$ (where the $cdf$ has bounded support). Details of the logic underlying equation~\ref{eqrstgen} are provided in \cite{fatma2023patient}. The expected remaining service time $w(x)$ estimated from equation~\ref{eqrstgen} can then be used in conjunction with equations~\ref{eq:lossub} and ~\ref{eq:wtpred} to estimate the RT-LOS at the subsystem under consideration.

This computation suffices for RT-LOS predictions at time $t$; however, the HFA algorithm (Algorithm \ref{alg:the_alg}) requires real-time LOS predictions at a future point in time $t + \delta$. 
Before we discuss RT-LOS prediction for individual PHC subsystems at future time $t + \delta$ based on the system state information available at current time $t$, we outline below the general approach for a general $M/G/m$ queuing system (procedure~\ref{deltaloscalc}).\\

\par\noindent\rule{\textwidth}{0.4pt}
\noindent \textit{Procedure 4.1.1: Compute the real-time length of stay for an $M/G/m$ queuing system at future time $t+\delta$ based on system state information available at current time $t$.}
\begin{enumerate}
\label{deltaloscalc}
    \item Collate system state information at time $t$: queue length $L_{q(t)}$, elapsed service times $x_{t(j)}$ of entities currently in service with each of the $j^{th}$ servers ($j = 1 \text{ to } m$).
    \item Estimate the expected system state at time $t + \delta$:
    \subitem[2.1] Estimate the expected number of new arrivals $A_{\delta}$ in $\delta$ amount of time. If $\lambda$ is the average interarrival time, then:
    \begin{equation}
        \label{eqarrgen}
        A_{\delta} = \max\left\{\frac{\delta}{\lambda} - 1,0\right\}.
    \end{equation}
    \subitem[2.2] Estimate the expected remaining service times $w(x_{t(j)}) ~ (j = 1 \text{ to } m)$  of the entities in service at the $m$ servers at time $t$ (using equations~\ref{eq:wgen} or ~\ref{eqrstgen}). The net remaining service time for the workstation will then be $w(x_{t}) = \min\{w(x_{t(1)}), w(x_{t(2)}), \dots, w(x_{t(m)})\}$. 
    
    \subitem[2.3] Estimate the expected number of entities who complete service in $\delta - w(x_t)$ amount of time, denoted by $N_{\delta}$. This is given by:
    \begin{equation}
    \label{eqdepgen}
        N_{\delta} = \min\left\{ L_{q(t)} + \frac{A_\delta}{2}, \sum\limits_{j = 1}^m \max \left\{\left\lfloor \frac{\delta - w(x_{t})}{E[X_j]} \right \rfloor, 0\right\}\right\}.
    \end{equation} 
    
    \subitem[2.4] Estimate the expected queue length at time $t +\delta$ as:
    \begin{equation}
        \label{eqlenqueuedelta}
        L_{q(t+\delta)} = L_{q(t)} + A_{\delta} - N_{\delta}.
    \end{equation}
    
    \subitem[2.5] Estimate the expected elapsed service time $x_{t+\delta(j)}$ of the entity in service with the $j^{th}$ server at time $t + \delta$, as given below:
    \begin{equation}
        \label{eqelapsedgen}
        x_{t+\delta(j)} = |\delta - w(x_{t(j)})| \mod E[X].
    \end{equation}

    \item Estimate the expected remaining service time of the workstation at $t+\delta$ as: 
    \begin{equation*}
        w(x_{t+\delta}) = \min\{w(x_{t+\delta(1)}), w(x_{t+\delta(2)}), \dots, w(x_{t+\delta(m)})\}.
    \end{equation*}
    
    \item Estimate the expected delay at time $t+\delta$ as:
    \begin{equation}
        \label{eqdeldeltagen}
        D_{t+\delta} = L_{q(t+\delta)} E[X] + w(x_{t+\delta})
    \end{equation}

    \item Compute the real-time LOS prediction at time $t + \delta$ as $L_{t +\delta} = D_{t+\delta} + E[X]$.
\end{enumerate}
\par\noindent\rule{\textwidth}{0.4pt}

The procedure above assumes that in general, there is an entity receiving service at the time the real-time LOS prediction is to be generated. If this is not the case, $x_t$ can simply be set to 0. 

Given that the real-time LOS prediction for $t + \delta$ is generated at $t$, system state information $s_t = (L_{q(t)}, x_t)$ is available only for $t$. The key step in procedure \ref{deltaloscalc} thus involves estimating the expected future system state $s_{t+\delta} = (L_{q(t+\delta)}, x_{t+\delta})$ as a function of $s_t = (L_{q(t)}, x_t)$. In order to compute $L_{q(t+\delta)}$, it can be seen from equation~\ref{eqlenqueuedelta} that the expected number of arrivals (equation \ref{eqarrgen}, where 1 is subtracted to account for the entity under consideration) and the expected number of entities completing service within $\delta$ amount of time must be estimated (equation~\ref{eqdepgen}). Computation of the number of entities completing service in $\delta$ time accounts for the expected remaining service time of the entity in service at time $t$ as well as the completion of the service of a few of the entities that arrive after $t$. This latter consideration is why $\frac{A_{\delta}}{2}$ is included in equation \ref{eqdepgen}. The elapsed service time of the entities in service at $t+\delta$, $x_{t+\delta}$, is computed from equation \ref{eqelapsedgen}, from which the expected remaining service time $w(x_{t+\delta})$ for the workstation is computed. $L_{q(t+\delta)}$ and $w(x_{t+\delta})$ are combined to calculate the expected real-time delay $D_{t+\delta}$ (equation \ref{eqdeldeltagen}), which when combined with the expected service time $E[X]$ yields the desired real-time LOS $L_{t+\delta}$. This procedure can be extended in similar fashion to $G/G/m$ queuing systems: for such systems, a computation similar to the procedure for computing expected remaining service time using the service process distributions will need to be done for the expected remaining time until arrival of the next entity. We now discuss the application of procedure~\ref{deltaloscalc} to the individual PHC subsystems, where applicable.

\noindent
\subsubsection{Real-time LOS prediction at the NCD nurse.}
\label{rtlosncd}

In computing the desired real-time LOS at the NCD nurse queuing system (subsystem $n \in I = \{n,d,l,p\}$), which we denote as $L_{t+\delta, n}$, we assume W.L.O.G. that a patient is undergoing service at $t$ with elapsed service time $x_{t,n}$ and that the queue length $L_{q(t, n)}$ is non-zero.


The NCD nurse's service time is modelled as a uniform distribution with limits $a, b~ (a < b)$ (see Table~\ref{tab1}). Applying equation~\ref{eqrstgen} for uniform service times, the expected remaining service time $w_{t, n}$, for the outpatient currently receiving service is estimated based on the elapsed service time $x_{t, n}$, as shown below. 

\begin{equation}
	\label{eqrstncd}
	w_{t, n} =
	\begin{cases}
		\frac{a+b}{2} - x_{t, n} ,        & 0 \le x_{t, n} < \frac{a+b}{2} \\
		\frac{a+3b}{4} - x_{t, n},       & \frac{a+b}{2} \le x_{t, n} < \frac{a+3b}{4} \\
		\frac{b - x_{t, n}}{2},          & \frac{a+3b}{4} \le x_{t, n}\le b \\	
	\end{cases}
\end{equation}


Per procedure~\ref{deltaloscalc}, we must first estimate $s_{t+\delta, n} = (L_{q(t+\delta, n)}, x_{t + \delta, n})$ as a function of $s_{t, n} = (L_{q(t, n)}, x_{t,n})$. Equations \ref{eqarrgen} and \ref{eqdepgen}, respectively, are used for this purpose with $\lambda$ from equation \ref{eqarrgen} set to $\lambda_n$, the average interarrival time of patients at subsystem $n$, and the number of servers $m$ set to 1. Here $\lambda_n = p_n \times \lambda_o$, where $p_n$ is the fraction of patients with age 30 years or greater, and $\lambda_o$ is the average outpatient interarrival time. For the remainder of the real-time LOS calculation for the NCD nurse queuing system, procedure \ref{deltaloscalc} is applied as described previously, yielding the estimate of expected real-time LOS at $t+\delta$, $L_{t+\delta, n}$.

\subsubsection{Real-time LOS prediction at the doctor.}
\label{rtlosdoc}

The doctor's service system is a multi-class queuing system, given that inpatients and childbirth patients also receive care from the doctor with non-preemptive priority over outpatients if they arrive during outpatient consultation hours. Hence, procedure~\ref{deltaloscalc} is not applicable directly to compute the real-time LOS for the outpatient at a future time point. However, as we will see shortly, the overarching approach remains similar, and portions of this procedure find use in the real-time LOS computation. 

From Algorithm~\ref{alg:loscalc}, it is seen that only patients aged 30 years or more visit the NCD nurse prior to consulting with the doctor. These patients thus reach the doctor's service system at a later time point (at $t_2 = t+\delta+L_n$) than patients aged less than 30 years (at $t_1 = t+ \delta$). Thus, we have to consider these two cases separately when computing the real-time LOS at the doctor's service system. We describe only the first case here, as the computation for the second case is very similar.   



In this scenario, we assume that the outpatient, after consulting with the NCD nurse, joins the queue (the outpatient department or OPD queue) for the doctor's queuing subsystem at time $t_{2} = t + \delta + L_{n}$. We assume W.L.O.G. that an outpatient is consulting with the doctor, with higher-priority inpatients and/or childbirth patients at the head of the queue, and previously arrived outpatients behind them. Thus the system state variables at time $t$ for generating the real-time LOS prediction for the outpatient are: (a) the elapsed service time of the outpatient in service with the doctor $x_{t,o}$, (b) the number of outpatients in the queue $N_{t,o}$, (c) the number of inpatients in the queue $N_{t,i}$, and (d) the number of childbirth patients in the queue $N_{t,c}$. Note that the inpatient and childbirth patient arrival processes are also modeled to be Poisson.

To estimate $L_{o(t_1)}$, the real-time LOS at the doctor's service system at $t_2$, we apply a procedure similar to \ref{deltaloscalc}: we consider the expected number of arrivals in $\delta + L_n$ time; the expected number of outpatient and higher-priority patients completing service by $t_2$; the expected remaining service time of the outpatient under service at $t$; and the elapsed service time of the patient receiving service at $t_2$. We now consider each of these.

The expected remaining service time for the outpatient currently under service with the doctor, $w_{t,o}$, is determined as a function of $x_{t, o}$, using equation~\ref{eqrstopd1}. Given that the Gaussian distribution ($\mathcal{N}(\mu, \sigma^2$)) models the outpatient's service process with the doctor (Table~\ref{tab1}), we set $G^{-1}{(0.5)} = \mu$, $G^{-1}{(0.75)} = \mu +0.675\sigma$, and $G^{-1}{(ext)} = \mu + 3\sigma$ in Equation~\ref{eqrstgen}.

\begin{equation}
	\label{eqrstopd1}
	w_{t, o} =
	\begin{cases}
		\mu - x_{t, o} ,        & 0 \le x_{t, o} < \mu \\
		(\mu + 0.675\sigma) - x_{t, o},       & \mu \le x_{t, o} < \mu + 0.675\sigma \\
		\frac{\mu + 3\sigma - x_{t, o}}{2},          & \mu + 0.675\sigma \le x_{t, o} \le \mu + 3\sigma \\	
	\end{cases}
\end{equation}




We first compute the average number of outpatients, inpatients, and childbirth patients arriving at the doctor's queue within $\delta + L_{n}$ time, denoted by $A_{t_2, o}$, $A_{t_2, i}$, and $A_{t_2, c}$ respectively. Each of the terms can be mathematically expressed as:

\begin{equation*}
	A_{{t_2, o}} = \max \left\{ \left (\frac{\delta + L_{n}}{\lambda_{o} \times (1 - p_{o, n})} + \max\left \{\left\lfloor \frac{\delta + L_n - w_{t,n}}{E[X_n]} \right\rfloor, 0 \right\} \right) - 1, 0 \right\};
\end{equation*}

\begin{equation*}
	A_{{t_2, i}} =  \frac{\delta + L_{n}}{\lambda_{i}};~ A_{{t_2, c}} = \frac{\delta + L_{n}}{\lambda_{c}}.
\end{equation*}


The expression for the number of outpatients arriving at the doctor's service system at $t_2$, $A_{t2, o}$, also accounts for the number of outpatients arriving from the NCD nurse's subsystem. Here $X_n$ is the service time of the NCD nurse. 


Owing to non-preemptive priority, higher-priority patients (inpatients and childbirth patients) are served immediately after the current outpatient. The average higher-priority patient service time, $\mu_{h}$, is computed as the weighted average of the average service times of the individual higher priority patient types, where the weights reflect the contribution of each individual higher-priority patient group to the total arrivals of higher-priority patients (equation~\ref{eqavgservicehp}).

\begin{equation}
	\label{eqavgservicehp}
	\mu_{h} = \lambda_{cbp}\times\frac{\mu_{cbp}}{\lambda_{cbp} + \lambda_{ip}} +  \lambda_{ip}\times\frac{\mu_{ip}}{\lambda_{cbp} + \lambda_{ip}}
\end{equation}

The expected number of higher-priority patients in the OPD queue at time ($\delta + L_{n} -  w_{t, o}$) will be given by equation~\ref{eqdelthp11}, while their elapsed service time $x_{t_{2}, h}$, expected remaining service time, $w_{t_2, h}$, and expected waiting time $d_{t_{2}, h}$ are computed as per the equations in Procedure~\ref{deltaloscalc}.


\begin{equation}
	\label{eqdelthp11}
	N_{t_{2}, h} = N_{t, i} + N_{t, c} + A_{t_{2}, i} + A_{t_{2}, c} - 	\max \left\{ \left\lfloor \frac{\delta + L_{n} - w_{t, o}}{\mu_{h}} \right\rfloor, 0 \right\}\\
\end{equation}





Now, the expected number of outpatients in the OPD queue at time $t_{2}$ and their corresponding wait time can be estimated as $N_{t_2, o} = N_{t, o} + A_{t_{2}, o};~ d_{t_2, o} = N_{t_2, o} \times E[X_{o}]$.






Once we have the closed-form expressions for $d_{t_{2}, h} $ and $d_{t_{2},o} $, the outpatient's overall `naive' delay at the doctor's queue at time $t_{2}$ will be the summation of delay due to higher-priority patients and the outpatients, which can be written as: $d_{opd} = d_{t_{2}, o}  + d_{t_2, h} \equiv d_{t_{2}, d_{(n)}}$. Here the computation of $d_{t_{2}, d_{(n)}}$ (`naively') assumes no new higher-priority arrivals during this period. However, one or more higher-priority patients may still arrive within the duration $d_{t_{2}, d_{(n)}}$ and move to the head of the queue. The expected number of such arrivals during $d_{t_{2}, d_{(n)}}$ is $N_{h} = \frac{d_{t_{2}, d_{(n)}}}{{\lambda_{h}}}$, where $\lambda_h$ ($\frac{1}{\lambda_{h}} = \frac{1}{\lambda_{ip}} + \frac{1}{\lambda_{cbp}}$) is the net interarrival time for higher-priority patients for priority classes $\lambda_{i}$ and $\lambda_{c}$ (\citet{kleinrock1965conservation}).




The delay caused by these potential arrivals is $d_{N_{h}} = N_{h} \mu_{h}  = \frac{d_{t_{2}, d_{(n)}}}{\lambda_{h}} \mu_{h}$.


Continuing this reasoning introduces additional opportunities for higher-priority patients to arrive, forming a geometric series for the total delay due to these patients:



\begin{equation*}
	d_{total} =  d_{t_{2}, d_{(n)}} + \frac{d_{t_{2}, d_{(n)}}}{\lambda_{h}} \mu_{h} + \frac{d_{t_{2}, d_{(n)}}}{\lambda_{h}^{2}} \mu_{h}^{2} + ...
\end{equation*}

Thus, the total delay at doctor's queue is 

\begin{equation}
	\label{delaytotal}
	d_{t_{2}, d} =  \frac{d_{t_{2}, d_{(n)}}}{ 1- \frac{\mu_{h}}{\lambda_{h}}} =  \frac{d_{t_{2}, d_{(n)}} \lambda_{h}}{ \lambda_{h} - \mu_{h}}.  \\
\end{equation}

Finally, $L_{d}$ at doctor's queue is $L_{d} =  d_{t_{2}, d} + E[X_{o}]$. A similar analysis is performed for the case where the outpatient joins the OPD queue without consulting the NCD nurse. We conclude by noting that the above computations assume that if higher-priority patients are present in the queue, then it is unlikely that service for these higher-priority patients is completed prior to the arrival of the outpatient under consideration at the facility. This is because for the specific facility that we consider, the average service times for the inpatients and childbirth patients with the doctor substantially exceeds that of the outpatient.

\subsubsection{Real-time LOS prediction at the laboratory PHC subsystem.}

\noindent Approximately 50\% of outpatients are routed to the in-house laboratory for routine investigations. To compute the real-time LOS in the laboratory queuing system at time $t_{3} = t + \delta + L_{n} + L_{o}$, denoted by $L_l$, we assume W.L.O.G. that an outpatient is currently in service with elapsed service time $x_{t, l}$, following prior consultations with NCD nurse and the doctor and the queue length is non-zero. The laboratory service time is modeled as a Gaussian distribution, and we use equation~\ref{eqrstopd1} to estimate the expected remaining service time, $w_{t, l}$. Per Procedure~\ref{deltaloscalc}, we compute the average number of outpatients arriving at the laboratory queue within $\delta + L_{n} + L_{o} $ time as $A_{t_3, l}$ while the expected number of patients in the laboratory queue is estimated using $N_{t_{3}, l}$.

\begin{equation}
	\label{eqdeltlabnew}
		A_{t_3, l} =  \min\left\{0.5 \times N_{t_{2}, o}, \left\lfloor\frac{L_o}{E[X_o]} \right\rfloor\right\}
\end{equation}


\begin{equation}
	\label{eqdellab}
	N_{t_{3}, l} = N_{t, l} + A_{t_3, l} - \max \left\{ \left\lfloor \frac{\delta + L_{n} + L{o} - w_{t, l}}{E[X_{l}]} \right\rfloor\ , 0 \right\} \\
\end{equation}

We leverage $N_{t_2, o}$ to estimate $A_{t_{3}, l}$, because it already accounts for the expected number of new outpatient arrivals within $\delta + L_n + L_o$ time and provides a more accurate estimate for the average number of arrivals, based on $\delta$ and $\lambda_{o}$. We then consider the number of patients who finish consultation with the doctor and reach the laboratory (with probability 0.5) by $t_3 = t+\delta + L_n + L_o$. Then, the elapsed service time at $t_{3}$, denoted by $x_{t_{3}, l}$, is estimated using equation~\ref{eqelapsedgen} in Procedure~\ref{deltaloscalc}.

\begin{equation}
	\label{eqndelopddellb}
	x_{t_{3}, l} = \mid (\delta + L_{n} + L_{o} - w_{t, l})\mid \text{~mod~} E[X_{l}] 
\end{equation}

The expected remaining service time $w_{t_{3}, l}$, and expected waiting time $d_{t_{3}, l}$ are computed via steps discussed in Procedure~\ref{deltaloscalc}. Finally, the real-time LOS at the laboratory service system is

\begin{equation}
	\label{lostotallab}
	L_{l} = d_{t_{3}, l} + E[X_{l}].
\end{equation}






A similar computation is performed for the case where outpatients directly visit the laboratory without visiting the NCD nurse.

\subsubsection{Real-Time LOS prediction at the pharmacy.}

\noindent To compute the real-time LOS at the pharmacy queuing system, denoted by $L_p$, at time $t_{4} =  t+ \delta + L_{n} + L_{o} + L_{l}$, we assume W.L.O.G. that an outpatient is currently in service with elapsed service time $x_{t, p}$ after completing consultations with the NCD nurse, doctor, and visiting the laboratory, and that the queue is non-empty. The pharmacy service time is modeled using a Gaussian distribution, and we use Equation~\ref{eqrstgen} to estimate the expected remaining service time $w_{t, p}$. The average number of outpatients arriving to the pharmacy queue $A_{t_{4}, p}$ and the effective queue length $N_{t_{4}, p}$ within ${\delta + L_{n} + L_{o} + L_{l}}$ time, is estimated as:

\begin{equation}
		\label{eqdeltpharmacy1}
		A_{t_{4}, p} =  \min \left \{N_{t_2, o} + N_{t,l}, \left \lfloor\frac{L_o}{E[X_o]}\right \rfloor + \left \lfloor\frac{L_l}{E[X_l]}\right\rfloor  \right\} 
\end{equation}

\begin{equation}
	\label{eqdelpharm}
	N_{t_{4}, p} = N_{t, p} + A_{t_{4}, p} - \max \left\{ \left\lfloor \frac{\delta + L_{n} + L_{o} + L_{l}  - w_{t, p}}{E[X_{p}]} \right\rfloor\, 0 \right\} \\
\end{equation}

The expected remaining service time for the outpatient, $w_{t_{4}, p}$, is estimated as a function of $x_{t_{4}, p}$ as per Procedure~\ref{deltaloscalc}. The expected waiting time, $d_{t_{4}, p}$, and the real-time LOS at $t_{4}$ at the pharmacy queue $L_{p}$ are then estimated similarly.

\begin{equation}
	\label{delaypharm1}
	\begin{aligned}
		&d_{t_{4}, p} = N_{t_{4}, p} \times E[X_{p}] + w_{t_{4}, p}, \text{ and } L_{p} =   d_{t_{4}, p} + E[X_{p}]
	\end{aligned}
\end{equation}

After estimating the expected LOS at each queuing subsystem, the total LOS for an outpatient is estimated per Algorithm~\ref{alg:loscalc}.



\subsection{\textit{Sim-ML Real-time LOS Prediction Approach for HFA}}
\label{secsimml}

As can be seen from the previous section, the development of the AQT RT-LOS predictor requires substantial effort even with relatively straightforward queuing systems such as those represented by patient flow in the PHC. We now describe the development of the Sim-ML RT-LOS predictor. The ease (or lack thereof) of developing the Sim-ML RT-LOS predictor) depends on whether the IT system required to capture the features required for the ML model that is part of the Sim-ML predictor can be or is already set up. As discussed before, such an IT system may simply be an extension of the billing enterprise resource planning (ERP) software system that is likely already to be present. For example, the time at which the patient arrives at the facility can be obtained as the time the patient checks in for registration at the facility reception, and the resource person (e.g., the doctor or the nurse) at each subsystem can simply log the time (perhaps at the click of a button on the ERP) at which service at their subsystem begins for a patient.    

The Sim-ML RT-LOS predictor development process can be summarized in Figure \ref{fig:simml} below.

\begin{figure}[htb]
    \centering
    \includegraphics[width=0.60 \textwidth]{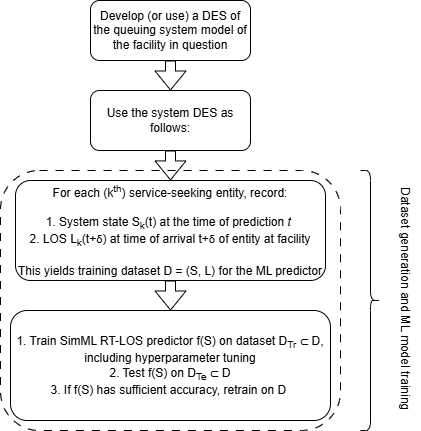}
    \caption{Sim-ML real-time length of stay predictor development process.}
    \label{fig:simml}
\end{figure}

We leverage the DES model of patient flow at the PHC to generate the input dataset for training Sim-ML regression models for generating RT-LOS estimates for outpatients. The simulation model records system state variables such as elapsed service time and queue length at each subsystem (NCD nurse, doctor, laboratory, pharmacy) relevant to the outpatient in the PHC. We also include features relating to the future system state at $t + \delta$ computed using the methods described in Section~\ref{secsimaqt}. Specifically, for each outpatient arrival, we record and/or compute the following variables.

\begin{enumerate}
    \item Queue lengths at time $t$ at each station.
    \item Expected remaining service times at time $t$ at each station (computed using methods described in Section~\ref{secsimaqt}).
    \item Travel times to each facility.
    \item Effective queue lengths at time $t+\delta$ at each station (computed).
    \item Expected remaining service times at time $t+\delta$ at each station (computed).
    \item Total time spent in the facility - i.e., the LOS, which forms the dependent variable or the label.
\end{enumerate}

The resulting dataset contains 32 system state features and a continuous target variable representing the patient's LOS in the facility. The final dataset contained 110,000 samples, which were obtained by running the simulation for 365 days following a 360-day warm-up period. A single replication required approximately 6 hours on an Intel \textit{i}7 64-bit Microsoft Windows operating system with 16 gigabytes of memory. Prior to the training process, we removed the outliers from the dataset using interquartile ranges computed with the help of box and whisker plots. For each ML model considered, the training set represented 75\% of observations, and the remaining belonged to the test set, which was used to evaluate the accuracy of each trained regressor. 

We trained various ML regressors, including the random forest, gradient-boosted, and XGBoost ensemble decision tree classifiers. Hyperparameter tuning was performed using the \textit{hyperopt} package in Python. For instance, the gradient boosting regressor used a learning rate of 0.012. We also trained a feed-forward neural network with three hidden layers using the Adam optimizer with \textit{tanh} activation function and a learning rate of 0.001. We provide additional information on model hyperparameters for each model in Table~\ref{tab_hyper} below.

\begin{table}[H]
	\begin{threeparttable}
		\centering
		\caption{Hyperparameters for the regressors.}
		\footnotesize
		\begin{tabular}{|l|p{10.455em}|l|}
			\hline
			\multicolumn{1}{|p{8.82em}|}{Regressors} & Hyperparameters & \multicolumn{1}{p{7.045em}|}{Estimates} \\
			\hline
			\multicolumn{1}{|l|}{{Random forest}} & max\_features & \multicolumn{1}{p{6.045em}|}{'sqrt'} \\
			& n\_estimators & 83 \\
			& n\_jobs & 1 \\
			\hline
			\multicolumn{1}{|l|}{{Gradient boosting }} & alpha & 0.94 \\
			& learning\_rate & 0.012 \\
			& max\_features & \multicolumn{1}{p{6.045em}|}{`sqrt'} \\
			& min\_samples\_leaf & 8 \\
			& n\_estimators & 364 \\
			& subsample & 0.96 \\
			\hline
			\multicolumn{1}{|l|}{{K-nearest neighbours}} & n\_jobs & 1 \\
			& n\_neighbors & 2 \\
			& metric & \multicolumn{1}{p{6.045em}|}{`manhattan'} \\
			\hline
			\multicolumn{1}{|l|}{{XGBoost}} & colsample\_bylevel & 0.61 \\
			& colsample\_bynode   & 1 \\
			& colsample\_bytree,  & 0.89 \\
			& gamma & 0.25 \\
			& learning\_rate & 0.08 \\
			& max\_depth & 2 \\
			& n\_estimators & 3800 \\
			& objective & \multicolumn{1}{p{6.045em}|}{`reg:linear'} \\
			& reg\_alpha & 0.006 \\
			& reg\_lambda  & 3.549 \\
			\hline
			\multicolumn{1}{|l|}{{Artificial neural network}} & hidden\_layer\_sizes & \multicolumn{1}{p{6.045em}|}{(8, 12, 20)} \\
			& activation & \multicolumn{1}{p{6.045em}|}{{`tanh'}} \\
			& solver & \multicolumn{1}{p{6.045em}|}{{`adam'}} \\
			& max\_iter & 500 \\
			& alpha & 0.0001 \\
			\hline
		\end{tabular}%
		\label{tab_hyper}%
	\end{threeparttable}
\end{table}%

We note here that the deployment and generation of these LOS predictions is unlikely to incur significant computational expense, regardless of the size of the facility network. This is because the generation of the LOS predictions at a facility involves just a single function evaluation, and this will remain the same regardless of the type of LOS predictor involved - for example, a queuing theory-based predictor or a statistical learning-based predictor. However, the specific LOS predictor to be employed must be selected taking into account the number of system state variables that are required for LOS prediction. For example, suppose a particular LOS predictor requires a large number of system state variables to be recorded and updated at high frequency. In that case, the complexity of the IT system required for this purpose will also increase. On the other hand, if another LOS predictor is available that requires significantly fewer variables to be recorded at a small loss of predictor accuracy, then it might be prudent to choose this latter LOS predictor from the standpoint of ease of deployment.


\subsection{Analysis of Sim-ML LOS Prediction Error}
\label{sec:simmlerror}

A natural question that can arise with respect to the Sim-ML approach is: how do simulation error and the ML regressor error interact and influence the final RT-LOS prediction? In the following discussion, we attempt to answer this question by setting up a simple analytical framework to analyze the Sim-ML predictor development and prediction process.

We first note that the Sim-ML prediction approach is likely to be adopted when sufficiently comprehensive historical data is not available to directly train an ML regressor. Therefore, we analyze Sim-ML error relative to the situation wherein sufficient historical data is available to train the ML regressor. 

Let an ML regressor $g(S|\beta)$ attempt to predict expected LOS $E[L] \in \mathbb{R}$ as a function of system state $S \in \mathbb{R}^d$ given a parameter set $\beta \in \mathbb{R}^p$. $\hat{\beta}_h$ is a set of parameter estimates obtained by fitting $g(.)$ on the historical dataset - e.g., a set of tuples $\{(S^h_1, L^h_1), (S^h_2, L^h_2), \dots, (S^h_m, L^h_m)\}$, yielding the predictor $g_h(S|\hat{\beta}_h)$. The $h$ subscript / superscript indicates the case where the ML LOS predictor is trained on historical data. We may assume that because the arrival of service-seeking entities to the facility is random, the system states recorded at the times of arrival of these entities are also random variables. Thus, each $(S^h_i, L^h_i)$ is a random tuple. 

Similarly, in the Sim-ML case, the dataset is generated via the DES of facility operations: e.g., a set of tuples $\{(S^v_1, L^v_1), (S^v_2, L^v_2), \dots, (S^v_m, L^v_m)\}$ (the $v$ subscript / superscript indicates the case where the ML LOS predictor is trained on synthetically or virtually generated data). This yields the ML regressor $g_v(S|\hat{\beta}_v)$ that, when combined with regression error, yields LOS prediction $L'$ as a function of system state $S$. This implies that, adopting a frequentist lens, both the historical data and the Sim-ML approaches attempt to estimate the `true' parameter set $\beta$ for the regressor. 

$\hat{L}$ and $L'$ are random variables, in that the predictors combine with regression error to yield the required predictions (equation~\ref{eq:regpred} below). 

\begin{equation}
    \label{eq:regpred}
    \begin{aligned}
        &\hat{L} = g_h(S|\hat{\beta}_h) + \epsilon_{rh}\\
        &L' = g_v(S|\hat{\beta}_v) + \epsilon_{rv}\\
    \end{aligned}
\end{equation}

In equation~\ref{eq:regpred}, we assume that the regression error random variable $\epsilon_r$ remains the same for both cases, with realizations $\epsilon_{rh}$ and $\epsilon_{rv}$ for the historical data and Sim-ML cases. This assumption appears reasonable given that we consider the case where the same regression model, with similar or the same hyperparameters, are used for both cases. Note that if we assume $E[\epsilon_r]$ = 0, then $g(S|\beta)$ can be considered to be an unbiased estimator of $E[L]$.

The use of the Sim-ML approach to train the RT-LOS ML regressor thus changes the parameter estimates of the regressor from $\hat{\beta}_h$ to $\hat{\beta}_v$. In other words, $\hat{\beta}_v = \hat{\beta}_h + \delta\beta$, where $\delta\beta$ represents the change in regressor parameter estimates when one adopts the Sim-ML approach instead of using historical data. This means we can also rewrite the expression for $L'$ as: 

\begin{equation*}
L' = g(S| \hat{\beta}_h + \delta\beta) + \epsilon_{rv}.  
\end{equation*}

Further, given that the same system state $S$ is given as input to both $g_h(.|\hat{\beta}_h)$ and $g_v(.|\hat{\beta}_v)$, we can drop the dependency on $S$ for the remainder of our analysis, and write: 

\begin{equation}
\label{eq:simmlbeta}
L' = g(\hat{\beta}_h + \delta\beta) + \epsilon_{rv}.  
\end{equation}

If the simulation is an adequate representation of the system, then $\delta\beta$ is likely to be small, and we can approximate equation~\ref{eq:simmlbeta} with its first-order Taylor series expansion around $g(\hat{\beta}_h)$. This is given below. 

\begin{equation}
	\label{eq:taylor}
	L' \approxeq g(\hat{\beta}_h) + (\delta\beta)^T\nabla g(\beta) + \epsilon_{rv}  
\end{equation}

Using the equation for $\hat{L}$ from equation~\ref{eq:regpred} in equation~\ref{eq:taylor}, we have:

\begin{equation}
	\label{eq:mlreg}
	L' \approxeq \hat{L} + (\delta\beta)^T\nabla g(\beta) + \epsilon_{rv} - \epsilon_{rh} 
\end{equation}

In equation~\ref{eq:mlreg}, the term $(\delta\beta)^T\nabla g(\beta)$ represents the constant error or bias, and the $\epsilon_{rv} - \epsilon_{rh}$ term represents the additional stochasticity introduced due to the use of the Sim-ML approach. We are now in a position to make the following statements about the Sim-ML approach relative to the historical data approach. 

\begin{proposition}
	\label{prop1}
	$E[L'] = E[\hat{L}] + (\delta\beta)^T \nabla g(\beta)$. Further, $Var(L') = Var(\hat{L}) + 2Var(\epsilon_r)$.
	\begin{proof}
		The first statement follows from the fact that $E[\epsilon_{rh}] = E[\epsilon_{rv}] = E[\epsilon_{r}]$ and the linearity of addition of expectations. The second statement follows from the fact that $\epsilon_{rh}$ and $\epsilon_{rv}$ are $iid$ realizations of $\epsilon_r$ and from the property of addition of variances. 
	\end{proof}
	
\end{proposition}

Now, we consider the implication of using a validated DES as part of the Sim-ML approach. We adopt the following definitions for what it means for a simulation to be validated in our analysis. 

\begin{definition}
	\label{defval1}
	For a discrete-event simulation with outcome $L'$ and real-world observations $\hat{L}$, the first level of validation implies that $E[L'] = E[\hat{L}]$. 
\end{definition}

The first level of validation implies that average values of outcomes estimated by the simulation are statistically comparable to those directly observed for the system in the `real world'. This level of validation is typically achieved by conducting hypothesis tests for the equality of expectations of the outcomes from the simulation versus those generated from the system in question. In our specific situation, this would involve the following condition. 

\begin{proposition}
	\label{prop2}
If the Sim-ML approach uses a DES that has achieved the first level of validation, then $(\delta\beta)^T\nabla g(\beta) = 0$.
	\begin{proof}
		The proof follows directly from Proposition~\ref{prop1}.   
	\end{proof}
\end{proposition}

Note that while additional levels of validation may be defined as involving the equality of variances and the distributions (more precisely, the $cdf$s) of $L'$ and $\hat{L}$, we do not consider these in our analysis given that the facility simulation in our case achieved only the first level of validation \citep{shoaib2022simulation}. Indeed, this is the case for most DESs of real-world system operations \citep{vonderheideextent}. 

\section{Computational Experimentation}
\label{rtlos_sec4}

In this section, we present the outcomes of computationally implementing the RT-HFA algorithm~\ref{alg:the_alg} across the two-facility healthcare facility network simulation. We begin by evaluating the AQT and Sim-ML LOS predictor accuracies. We then present the effective arrival rates for outpatients observed at the two PHCs at steady state. Finally, we compare operational outcomes at the two PHCs under four distinct scenarios: (a) the status quo (no RT-HFA); (b) RT-HFA with actual real-time LOS estimates, (c) RT-HFA with RT-LOS predictions from the AQT predictor, and (d) RT-HFA with RT-LOS predictions from the Sim-ML predictor.

We quantified the accuracy of both predictors using the mean absolute percentage error (MAPE) (see Table~\ref{tab:combined_mape}), computed as $\text{MAPE} = \frac{1}{N}\sum\limits_{i=1}^{N}\left|\frac{LOS_{a, i}-LOS_{p, i}}{LOS_{a, i}}\right|$. Here {LOS}$_{a, i}$ denotes the actual RT-LOS of the i$^{th}$ patient, {LOS}$_{p, i}$ denotes the predicted RT-LOS of the i$^{th}$ patient, and $N$ is the total number of patients (predictions) in the sample.

We report RT-LOS predictor accuracy across two settings. First, we consider RT-LOS predictor accuracy when considering patient flow though the PHC (recall Figure \ref{Fig1a}): (i) patients visit the NCD nurse, the doctor and the pharmacy; (ii) patient flow in which all consultations are routed through the NCD nurse; and (iii) patient flow without NCD nurse consultation, but with lab visits. Second, we report LOS predictor accuracy at each queuing subsystem: (i) NCD nurse, (ii) doctor, (iii) laboratory, and (iv) pharmacy. For each predictor, we report the mean and standard deviation (SD) of the MAPE scores across 15 replications.

Table~\ref{tab:combined_mape} highlights significant variability in the performance of the Sim-ML predictors. Among them, the XGBoost (XGB), K-Nearest Neighbors (KNN), and Artificial Neural Network (ANN) models achieve the best performance in the flow-wise setting, which is the relevant use-case from an RT-HFA perspective. The ANN model, in particular, yields the lowest MAPE scores, ranging from 2.33\% to 3.17\%, indicating encouraging capability in capturing LOS variations across different patient flows. The Random Forest model exhibits reasonable performance but exhibits greater variability with higher MAPE scores compared to other predictors. The AQT predictor, while less precise than the best Sim-ML predictors, demonstrates stable performance, with MAPE scores remaining close to 9\%.

\begin{table}[htbp]
    \centering
    \caption{Real-time length of stay predictor accuracies.}
    \label{tab:placeholder_label}
    \begin{tabular}{|p{4.5em}|p{5.1em}|p{5.2em}|p{5.2em}|p{5.2em}|p{5.2em}|p{5.2em}|}
        \hline
        Cases & AQT Predictor & \multicolumn{5}{|c|}{Sim-ML} \\
        \cline{3-7}
         &  & Random forest & Gradient-boosted trees & KNN & XGB & ANN \\
        \hline
        \multicolumn{7}{|c|}{Flow-wise RT-LOS predictor MAPE values} \\
        \hline
        Case 1 & 8.84 (0.21) & 26.22 (0.30) & 6.71 (0.36) & 7.65 (0.14) & 6.12 (0.34) & 3.17 (0.14) \\
        Case 2 & 8.90 (0.08) & 17.53 (0.09) & 7.16 (0.13) & 5.80 (0.08) & 3.07 (0.12) & 3.43 (0.12) \\
        Case 3 & 9.04 (0.18) & 25.17 (0.22) & 7.52 (0.38) & 6.37 (0.10) & 5.05 (0.11) & 2.33 (0.06) \\
        \hline
        \multicolumn{7}{|c|}{PHC subsystem-wise RT-LOS predictor MAPE values} \\
        \hline
        NCD        & 5.55 (0.08) & 10.51 (0.19) & 6.21 (0.09) & 1.74 (0.05) & 28.10 (0.35) & 11.01 (0.19) \\
        OPD        & 17.55 (0.24) & 23.44 (0.10) & 23.42 (0.09) & 22.64 (0.25) & 14.01 (0.22) & 21.60 (0.22) \\
        Laboratory & 19.01 (0.11) & 23.13 (0.74) & 14.12 (0.19) & 24.62 (0.44) & 13.54 (0.20) & 11.34 (0.19) \\
        Pharmacy   & 9.01 (0.12) & 23.95 (0.19) & 22.31 (0.25) & 23.12 (0.21) & 9.11 (0.21) & 21.17 (0.11) \\
        \hline
    \end{tabular}
    \smallskip
\footnotesize
Case 1: NCD$\to$OPD$\to$Pharmacy; Case 2: NCD$\to$OPD$\to$Lab$\to$Pharmacy; Case 3: OPD$\to$Lab$\to$Pharmacy. \\
AQT: analytical queuing-theoretic; KNN: K-nearest neighbor; XGB: Extreme gradient-boosted trees; ANN: artificial neural network; PHC: primary health center; MAPE: mean absolute percentage error.

\label{tab:combined_mape}
\end{table}


With respect to subsystem-wise flow, KNN achieves the lowest MAPE (1.74\%) at the NCD station, whereas XGB exhibits relatively high error. Both AQT and Sim-ML predictors show elevated MAPE scores at the OPD station, with ANN accuracy at 21.60\%. 
KNN, XGB, and ANN demonstrate comparatively better accuracy at the laboratory and pharmacy stations; that said, MAPE scores remain relatively higher overall, likely due to the complexity of interactions at these stations. However, it appears that relatively lower performances (MAPE scores exceeding 10\%) of RT-LOS predictors at individual PHC subsystems combine via antithetical error combinations to yield better overall flow-wise performance.  

In summary, among Sim-ML predictors, the ANN and KNN models perform well across both flow-wise and station-wise analyses; their effectiveness varies by station. Overall, it is the AQT predictor that exhibits the most stable performance across both flow-wise and subsystem settings. This is likely because the operational dynamics in these (sub)systems are captured reasonably well by the AQT predictor. The Sim-ML approach, on the other hand, is advantageous in that it can be applied to generate RT-LOS (or indeed real-time delay) predictions regardless of the underlying queuing system dynamics. That said, systems with substantial nonstationarity may require a more complex training and prediction process (e.g., different models for different phases).

To compare the operational outcomes for evaluating the performance of implementing the RT-HFA algorithm across the facility network, we programmed the simulation in the Python programming language, using the \textit{salabim} package within the IntelliJ IDEA integrated environment. All experiments were conducted on a workstation equipped with an Intel \textit{i}7 processor, running on a 64-bit Microsoft Windows operating system with 16 gigabytes of memory. The simulation was run for 365 days of PHC operations, with an additional 360 day warm-up period, and outcomes were generated from 40 independent replications. The total runtime required to generate datasets from 40 replications was approximately 120 minutes.

Before describing outcomes from the computational implementation of the RT-HFA algorithm~\ref{alg:the_alg}, we note that when any HFA method is implemented, the arrival rate at each facility in the network changes after implementation. This is relevant for both the AQT and Sim-ML predictors, as they use the expected arrival rate in future system state calculations (for the Sim-ML predictor, this is applicable for the engineered system state features for $t + \delta$). Therefore, it is important to compute the effective arrival rate of outpatients at each healthcare facility, $\lambda_{eff}$, realized after implementation of the RT-HFA method. Algorithm~\ref{alg:leff} outlines this procedure.

\begin{algorithm}
	\caption{Effective average interarrival time estimation upon RT-HFA implementation.}
	\label{alg:leff}
	\begin{algorithmic}[1]
	\State Initialize threshold $T_0$, `original' average interarrival time $\lambda$, counter $i = 0$, termination threshold $\epsilon$.
	\State $\lambda_{i} = M \times \lambda $ \Comment{arbitrary initialization of $\lambda_i$ to ensure at least one execution of the loop}
	\While{$\lambda_{i} - \lambda < \epsilon$} 
		\State Implement RT-HFA with $\lambda$
		\For{$t = T_0 \times i$ to $T_0 \times (i+ 1)$}
			\State Record number of outpatient arrivals $N$ in $T_0$ to the facility under consideration
			\State {$\lambda_{i} = \frac{T_0}{N}$}
			\State $\lambda = \lambda_{i}$
			\State $i \gets i + 1$
		\EndFor
        \State $\lambda_{\text{eff}} = \lambda_{i}$
	\EndWhile
	\end{algorithmic}
\end{algorithm}

Algorithm~\ref{alg:leff} involves implementing the RT-HFA algorithm using the initial interarrival time of outpatients at each facility within the network for a threshold duration of $T_o$ days, after which the average interarrival time at each facility is updated based on the net number of outpatients arriving at each facility in the network after implementation of RT-HFA. The RT-HFA algorithm is then re-implemented with the updated average interarrival time until $\lambda_{eff}$ stabilizes. We describe the computational implementation of Algorithm~\ref{alg:leff} to estimate $\lambda_{eff}$ in Section~\ref{rtlos_sec5}.

\begin{figure}[!ht]
	\centering
	\begin{subfigure}{0.46\textwidth}
		\includegraphics[width=\linewidth]{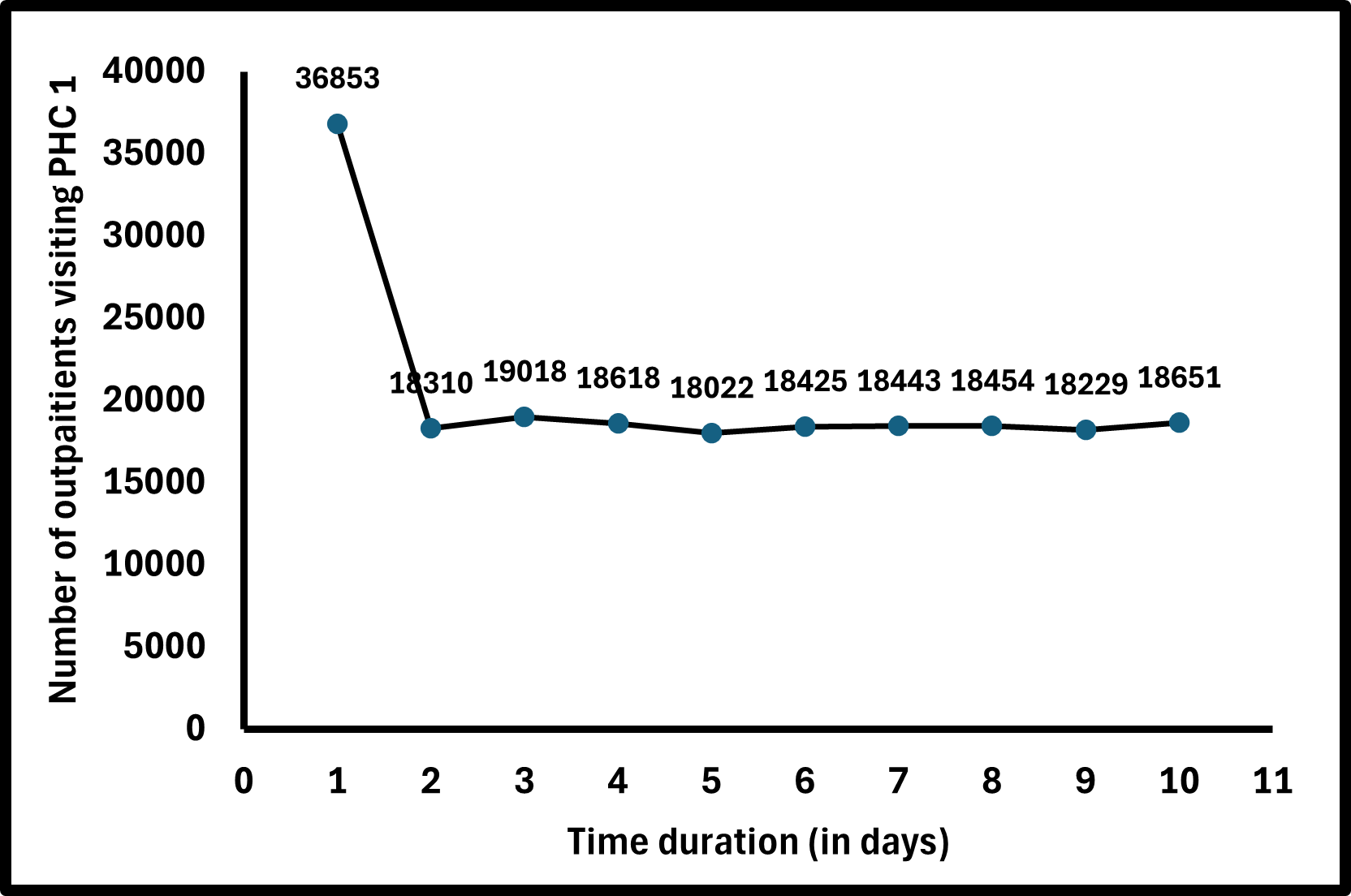}
		\caption{$\lambda_{eff}$ at PHC 1, estimated using actual RT-LOS.}
	\end{subfigure}
	\hspace{20pt}
	\begin{subfigure}{0.46\textwidth}
		\includegraphics[width=\linewidth]{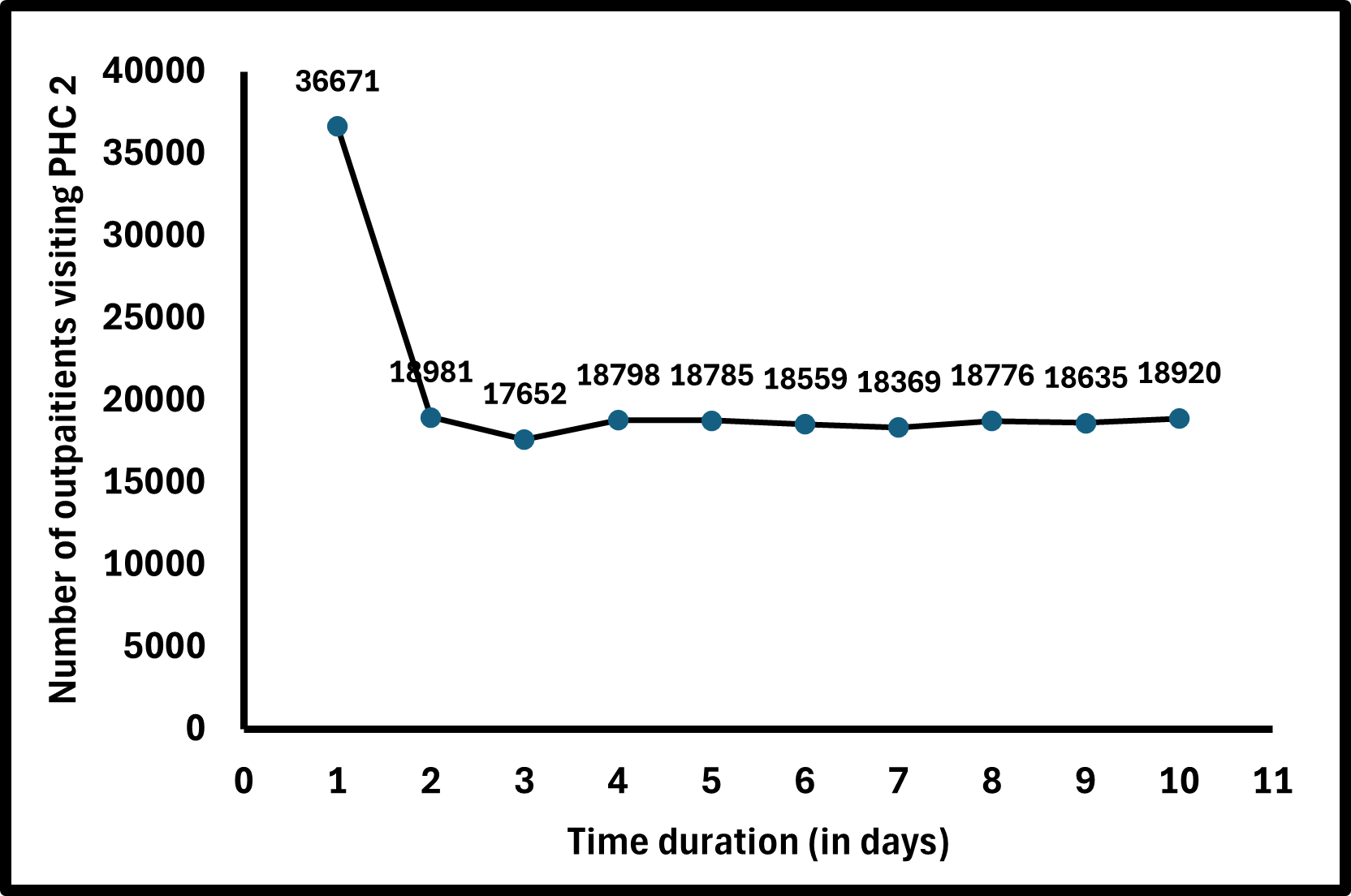}
		\caption{$\lambda_{eff}$ at PHC 2, estimated using actual RT-LOS.}
	\end{subfigure}
	\begin{subfigure}{0.46\textwidth}
		\includegraphics[width=\linewidth]{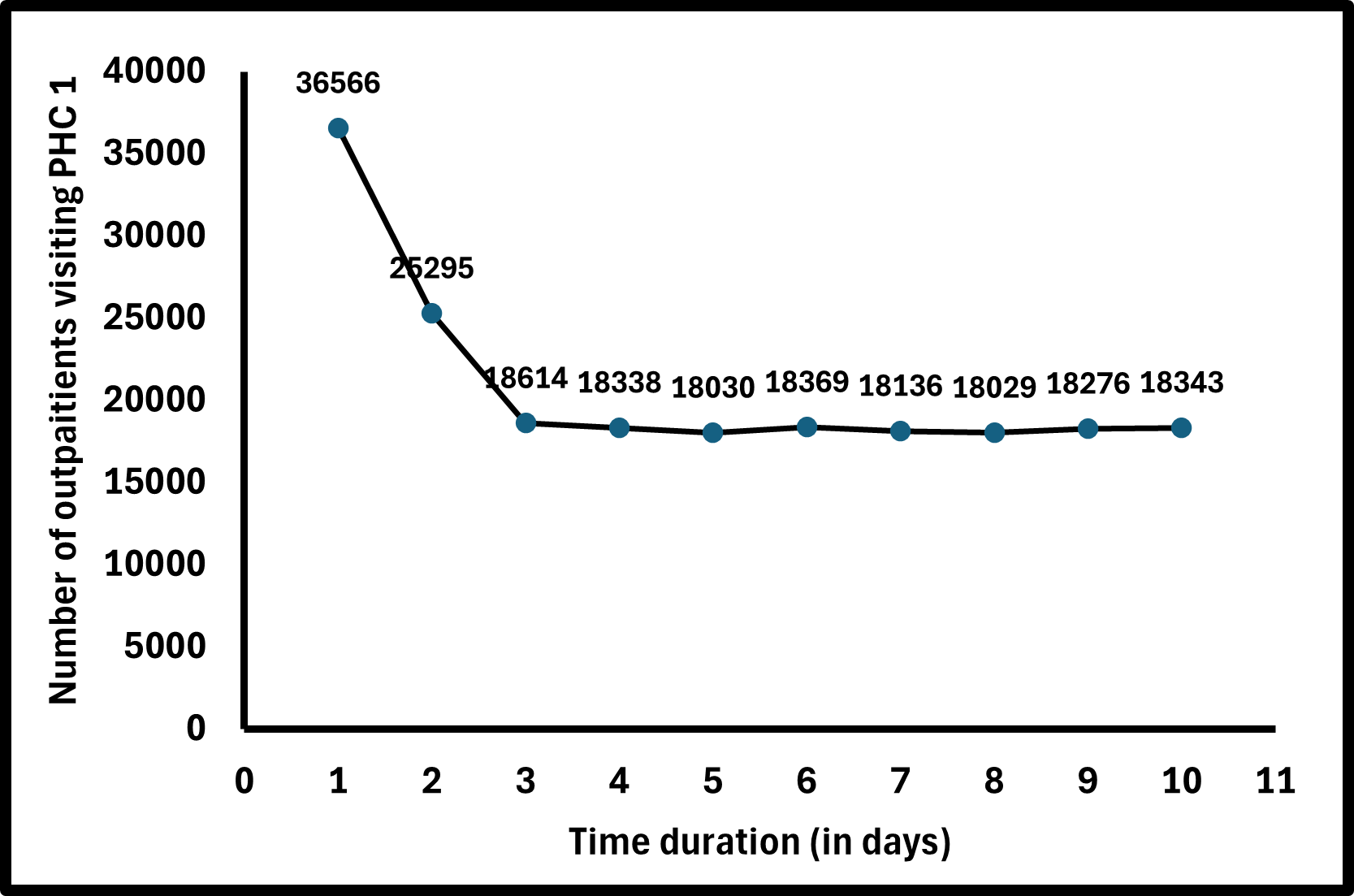}
		\caption{$\lambda_{eff}$ at PHC 1, estimated using predicted RT-LOS.}
	\end{subfigure}
	\hspace{20pt}
	\begin{subfigure}{0.46\textwidth}
		\includegraphics[width=\linewidth]{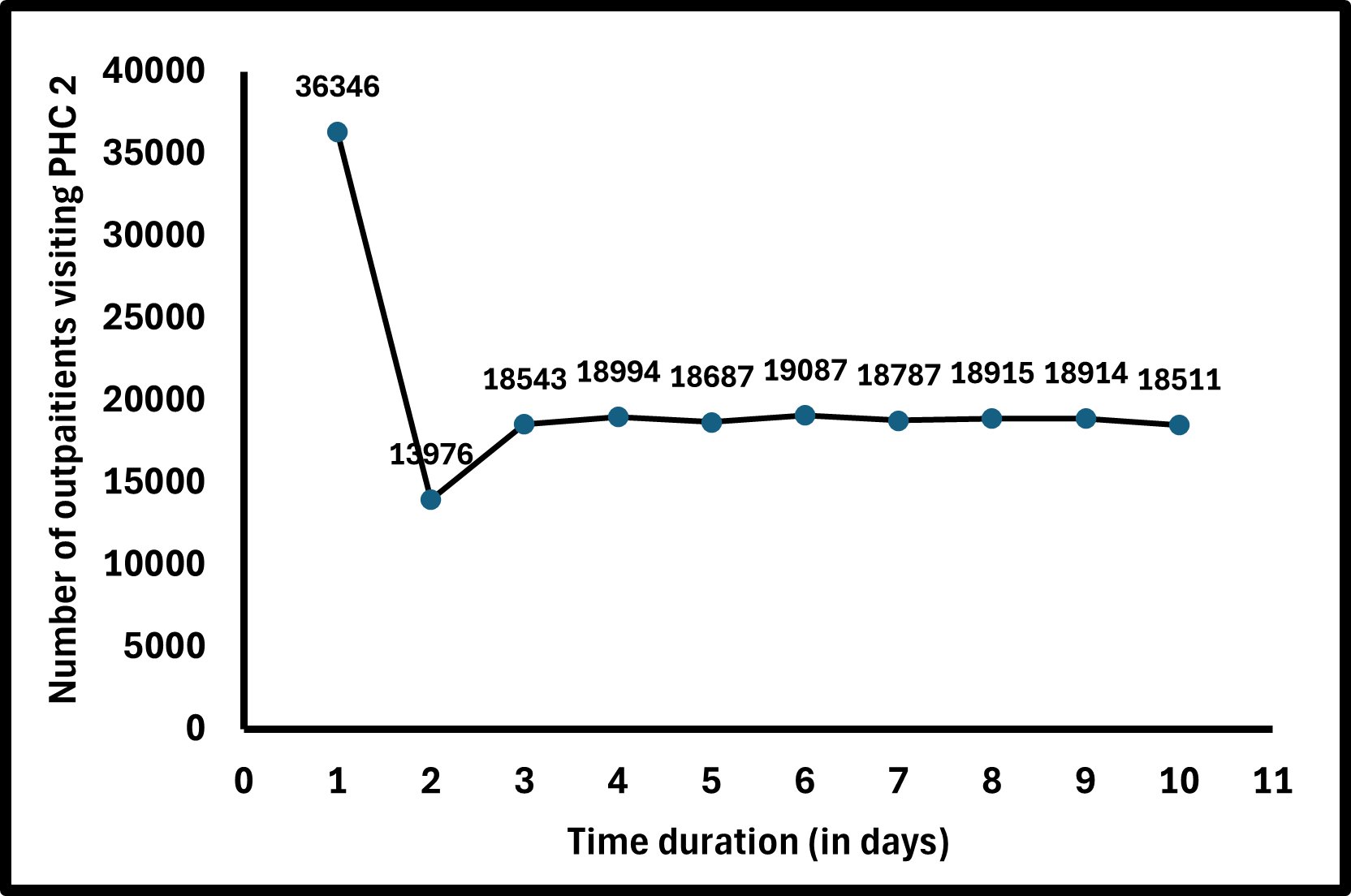}
		\caption{$\lambda_{eff}$ at PHC 2, estimated using predicted RT-LOS.}
	\end{subfigure}
	
	\caption{Variation in the number of outpatients and estimation of steady-state effecttive average interarrival time post RT-HFA implementation.}
	\label{fig:steadystate}
\end{figure}

We first recorded the average number of outpatients visiting the PHCs when the RT-HFA algorithm was implemented using both actual and predicted LOS estimates. This involved implementing Algorithm~\ref{alg:leff} to determine the effective post-RT-HFA outpatient arrival rate.  Figure~\ref{fig:steadystate} presents the results for PHCs 1 and 2. The results indicate that outpatient visit volumes stabilize after approximately 360 days for both cases (with each time unit in the figure corresponding to 90 days). Once steady state was reached, we began collecting network-level operational outcomes from the RT-HFA implementation and generated the datasets to train the Sim-ML predictors. This steady-state analysis also formed the basis for the selection of an appropriate warm-up period for the simulation. 

We now present the simulation results generated from implementing the RT-HFA algorithm across PHC$_{1}$ and PHC$_{2}$ in Table~\ref{table2f}. For all four scenarios, we report the mean and SD of key operational outcomes, including resource occupancies, wait times, and average lengths of stay. We also define an important metric, $\Delta_{net(o)}$, which represents the percentage difference between the maximum and minimum value of operational outcome $o$ across the healthcare facilities in the network, evaluated as:

\begin{equation}
\label{eqqui}
\Delta_{\text{net(o)}}
=
\left|
\frac{\max_{o} - \min_{o}}{\max_{o}}
\right|
\times 100.
\end{equation}

For example, for the doctor's utilization, the notation is modified as $\Delta_{net(\rho-doc)}$, where $\rho_{doc}$ denotes utilization of the doctor. In general, $\Delta_{net(o)}$ measures the extent to which that outcome $o$ is equitably distributed across the network. Lower values of $\Delta_{net(o)}$ indicate more equitable use of resources across the network. However, $\Delta_{net(o)}$ should not be interpreted in isolation when evaluating the effectiveness of an assignment algorithm. An assignment algorithm may reduce $\Delta_{net(o)}$ relative to the no-assignment scenario, indicating improved resource utilization balance and patient distribution across facilities, but may still result in average lengths of stay that exceeds clinically meaningful thresholds. In such cases, the assignment algorithm alone may be insufficient to alleviate congestion. Therefore, $\Delta_{net(o)}$ must be interpreted alongside other operationally and clinically relevant performance measures.

We also report the percentage of patients who are diverted, defined as the proportion of outpatients who are assigned to a PHC different from their usual choice (their nearest facility) under the RT-HFA algorithm. This metric allows us to assess, with the incorporation of compliance mechanisms, how many of the patients who would normally visit PHC$_1$ are now diverted to PHC$_2$, and vice versa, because their expected LOS is estimated to be shorter at the alternative facility, as determined by Algorithm~\ref{alg:the_alg}.

The outcomes reported in Table~\ref{table2f} highlight substantial improvements in operational performance achieved by assigning healthcare facilities per Algorithm~\ref{alg:the_alg}, with more equitable distribution of patient load and service outcomes across both PHCs. In the baseline scenario without RT-HFA, patients on average spent substantially higher LOS (58.492 minutes) at PHC$_2$ compared to only 8.582 minutes at PHC$_1$. However, with the assignment implemented using actual real-time LOS estimates, the average LOS at PHC$_2$ decreased to 9.674 minutes, while the LOS at PHC$_1$ increased slightly to 9.494 minutes, resulting in a substantially balanced distribution. When the assignment algorithm was implemented using the AQT and Sim-ML predictor, the same trend was observed, with better (lower) LOS values observed at PHC$_2$. The corresponding $\Delta_{net(o)}$ values for LOS also reduced substantially: 1.86\%, 6.31\%, 10.58\% for RT-HFA with actual, AQT, Sim-ML predictors, respectively, compared to 85.32\% in the scenario without RT-HFA implementation.

\begin{table}[H]
	\begin{threeparttable}
    \footnotesize
	\centering
	\caption{Operational outcomes from simulating the implementing the real-time healthcare facility assignment algorithm across a two-PHC network.}
	\begin{tabular}{|R{6cm}|p{4.045em}|p{9.635em}ll|}
		\hline
		\multicolumn{1}{|p{4.775em}|}{Cases} & \multicolumn{1}{|p{5em}|}{Outcomes*} & \multicolumn{1}{p{6em}|}{PHC$_1$ \newline{9/1/1/1/1/1/1}} & \multicolumn{1}{p{6em}|}{PHC$_2$ \newline{2/1/1/1/1/1/1}} & \multicolumn{1}{p{6.725em}|}{$\Delta_{net(o)}$} \\
		\hline
		{Without RT-HFA } &  ${\rho_{doc}}$     & \multicolumn{1}{p{5.635em}|}{0.463 (0.003)} & \multicolumn{1}{p{6em}|}{1.085 (0.004)} & \multicolumn{1}{p{6.725em}|}{57.31\% (0.38)} \\
		\cline{2-5}         &  ${\rho_{ncd}}$     & \multicolumn{1}{p{5.635em}|}{0.513 (0.005)} & \multicolumn{1}{p{6em}|}{1.472 (0.017)} & \multicolumn{1}{p{6.725em}|}{65.11\% (0.45)} \\
		\cline{2-5}          &  ${\rho_{phar}}$     & \multicolumn{1}{p{5.635em}|}{0.383 (0.003)} & \multicolumn{1}{p{6em}|}{1.704 (0.006)} & \multicolumn{1}{p{6.725em}|}{77.52\% (0.22)} \\
		\cline{2-5}          &  ${\rho_{lab}}$     & \multicolumn{1}{p{5.635em}|}{0.317 (0.003)} & \multicolumn{1}{p{6em}|}{1.414 (0.006)} & \multicolumn{1}{p{6.725em}|}{77.56\% (0.25)} \\
		\cline{2-5}          &  ${w_{opd}}$     & \multicolumn{1}{p{5.635em}|}{0.367 (0.081)} & \multicolumn{1}{p{6em}|}{2.843 (0.185)} & \multicolumn{1}{p{6.725em}|}{51.68\% (4.61)} \\
		\cline{2-5}          &  ${w_{phar}}$    & \multicolumn{1}{p{5.635em}|}{0.353 (0.013)} & \multicolumn{1}{p{6em}|}{35.82 (0.729)} & \multicolumn{1}{p{6.725em}|}{99.01\% (0.04)} \\
		\cline{2-5}          &  ${w_{lab}}$     & \multicolumn{1}{p{5.635em}|}{0.522 (0.010)} & \multicolumn{1}{p{6em}|}{18.58 (0.599)} & \multicolumn{1}{p{6.725em}|}{97.18\% (0.001)} \\
		\cline{2-5}          &   ${w_{ncd}}$     & \multicolumn{1}{p{5.635em}|}{0.434 (0.018)} & \multicolumn{1}{p{6em}|}{9.417 (0.387)} & \multicolumn{1}{p{6.725em}|}{95.37\% (0.27)} \\
		\cline{2-5}          &  ${LOS}$     & \multicolumn{1}{p{5.635em}|}{8.582 (0.123)} & \multicolumn{1}{p{6em}|}{58.492 (1.20)} & \multicolumn{1}{p{6.725em}|}{85.32\% (0.39)} \\
		\cline{2-5}          &  $\beta^{*}$    & \multicolumn{3}{p{16.635em}|}{0.00(0.00)} \\
		\hline
		{With RT-HFA (actual LOS)} & ${\rho_{doc}}$     & \multicolumn{1}{p{5.635em}|}{0.767 (0.002)} & \multicolumn{1}{p{6em}|}{0.770 (0.003)} & \multicolumn{1}{p{6.725em}|}{0.41\% (0.002)} \\
		\cline{2-5}         & ${\rho_{ncd}}$    & \multicolumn{1}{p{5.635em}|}{0.979 (0.005)} & \multicolumn{1}{p{6em}|}{0.981 (0.006)} & \multicolumn{1}{p{6.725em}|}{0.48\% (0.004)} \\
		\cline{2-5}          & ${\rho_{phar}}$    & \multicolumn{1}{p{5.635em}|}{1.038 (0.003)} & \multicolumn{1}{p{6em}|}{1.049 (0.003)} & \multicolumn{1}{p{6.725em}|}{1.01\% (0.003)} \\
		\cline{2-5}          &  ${\rho_{lab}}$    & \multicolumn{1}{p{5.635em}|}{0.801 (0.004)} & \multicolumn{1}{p{6em}|}{0.870 (0.003)} & \multicolumn{1}{p{6.725em}|}{7.94\% (0.004)} \\
		\cline{2-5}          &  ${w_{opd}}$     & \multicolumn{1}{p{5.635em}|}{0.646 (0.020)} & \multicolumn{1}{p{6em}|}{0.681 (0.020)} & \multicolumn{1}{p{6.725em}|}{5.36\% (0.031)} \\
		\cline{2-5}          &  ${w_{phar}}$     & \multicolumn{1}{p{5.635em}|}{1.308 (0.039)} & \multicolumn{1}{p{6em}|}{1.405 (0.073)} & \multicolumn{1}{p{6.725em}|}{7.21\% (0.048)} \\
		\cline{2-5}          &   ${w_{lab}}$     & \multicolumn{1}{p{5.635em}|}{1.255 (0.066)} & \multicolumn{1}{p{6em}|}{1.470 (0.030)} & \multicolumn{1}{p{6.725em}|}{14.63\% (0.037)} \\
		\cline{2-5}          &   ${w_{ncd}}$     & \multicolumn{1}{p{5.635em}|}{1.182 (0.020)} & \multicolumn{1}{p{6em}|}{1.195 (0.038)} & \multicolumn{1}{p{6.725em}|}{2.77\% (0.021)} \\
		\cline{2-5}          &   ${LOS}$     & \multicolumn{1}{p{5.635em}|}{9.494 (0.101)} & \multicolumn{1}{p{6em}|}{9.674 (0.062)} & \multicolumn{1}{p{6.725em}|}{1.86\% (0.006)} \\
		\cline{2-5}          &  $\beta^{*}$     & \multicolumn{3}{p{16.635em}|}{49.51 (0.18)} \\
		\hline
		{With RT-HFA: AQT predictor} & ${\rho_{doc}}$     & \multicolumn{1}{p{5.635em}|}{0.760 (0.003)} & \multicolumn{1}{p{6em}|}{0.780 (0.003)} & \multicolumn{1}{p{6.725em}|}{2.54\% (0.005)} \\
		\cline{2-5}          & ${\rho_{ncd}}$     & \multicolumn{1}{p{5.635em}|}{0.962 (0.007)} & \multicolumn{1}{p{6em}|}{1.002 (0.007)} & \multicolumn{1}{p{6.725em}|}{3.97\% (0.009)} \\
		\cline{2-5}          & ${\rho_{phar}}$     & \multicolumn{1}{p{5.635em}|}{1.014 (0.006)} & \multicolumn{1}{p{6em}|}{1.074 (0.005)} & \multicolumn{1}{p{6.725em}|}{5.55\% (0.009)} \\
		\cline{2-5}          &  ${\rho_{lab}}$    & \multicolumn{1}{p{5.635em}|}{0.785 (0.005)} & \multicolumn{1}{p{6em}|}{0.889 (0.005)} & \multicolumn{1}{p{6.725em}|}{11.70\% (0.009)} \\
		\cline{2-5}          & ${w_{opd}}$     & \multicolumn{1}{p{5.635em}|}{0.601 (0.029)} & \multicolumn{1}{p{6em}|}{0.690 (0.021)} & \multicolumn{1}{p{6.725em}|}{12.80\% (0.048)} \\
		\cline{2-5}          &  ${w_{phar}}$     & \multicolumn{1}{p{5.635em}|}{1.289 (0.027)} & \multicolumn{1}{p{6em}|}{1.494 (0.038)} & \multicolumn{1}{p{6.725em}|}{8.67\% (0.047)} \\
		\cline{2-5}          &   ${w_{lab}}$     & \multicolumn{1}{p{5.635em}|}{1.194 (0.024)} & \multicolumn{1}{p{6em}|}{1.589 (0.038)} & \multicolumn{1}{p{6.725em}|}{24.89\% (0.020)} \\
		\cline{2-5}          &  ${w_{ncd}}$    & \multicolumn{1}{p{5.635em}|}{1.055 (0.024)} & \multicolumn{1}{p{6em}|}{1.270 (0.049)} & \multicolumn{1}{p{6.725em}|}{16.80\% (0.032)} \\
		\cline{2-5}          &  ${LOS}$    & \multicolumn{1}{p{5.635em}|}{9.105 (0.058)} & \multicolumn{1}{p{6.5em}|}{10.200 (0.151)} & \multicolumn{1}{p{6.725em}|}{6.31\% (0.015)} \\
		\cline{2-5}          &  $\beta^{*}$    & \multicolumn{3}{p{16.635em}|}{47.98 (0.13)} \\
		\hline
		{With RT-HFA: Sim-ML predictor (KNN model)} & ${\rho_{doc}}$     & \multicolumn{1}{p{5.635em}|}{0.754 (0.004)} & \multicolumn{1}{p{6em}|}{0.785 (0.003)} & \multicolumn{1}{p{6.725em}|}{3.93\%(0.00)} \\
\cline{2-5}    & ${\rho_{ncd}}$     & \multicolumn{1}{p{5.635em}|}{0.958 (0.006)} & \multicolumn{1}{p{6em}|}{1.006 (0.007)} & \multicolumn{1}{p{6.725em}|}{4.81\% (0.00)} \\
\cline{2-5}          & ${\rho_{phar}}$     & \multicolumn{1}{p{5.635em}|}{1.003 (0.006)} & \multicolumn{1}{p{6em}|}{1.083 (0.006)} & \multicolumn{1}{p{6.725em}|}{7.41\% (0.01)} \\
\cline{2-5}          &  ${\rho_{lab}}$    & \multicolumn{1}{p{5.635em}|}{0.777 (0.005)} & \multicolumn{1}{p{6em}|}{0.900 (0.006)} & \multicolumn{1}{p{6.725em}|}{13.63\% (0.01)} \\
\cline{2-5}          & ${w_{opd}}$     & \multicolumn{1}{p{5.635em}|}{0.600 (0.026)} & \multicolumn{1}{p{6em}|}{0.708 (0.023)} & \multicolumn{1}{p{6.725em}|}{15.15\% (0.04)} \\
\cline{2-5}          &  ${w_{phar}}$     & \multicolumn{1}{p{5.635em}|}{1.282 (0.066)} & \multicolumn{1}{p{6em}|}{1.502 (0.063)} & \multicolumn{1}{p{6.725em}|}{14.57\% (0.03)} \\
\cline{2-5}          &   ${w_{lab}}$     & \multicolumn{1}{p{5.635em}|}{1.188 (0.084)} & \multicolumn{1}{p{6em}|}{1.625 (0.053)} & \multicolumn{1}{p{6.725em}|}{26.80\% (0.06)} \\
\cline{2-5}          &  ${w_{ncd}}$    & \multicolumn{1}{p{5.635em}|}{1.042 (0.068)} & \multicolumn{1}{p{6em}|}{1.299 (0.041)} & \multicolumn{1}{p{6.725em}|}{17.84\% (0.06)} \\
\cline{2-5}          &  ${LOS}$    & \multicolumn{1}{p{5.635em}|}{9.097 (0.130)} & \multicolumn{1}{p{6.5em}|}{10.321 (0.178)} & \multicolumn{1}{p{6.725em}|}{10.58\% (0.01)} \\
\cline{2-5}          &  $\beta^{*}$    & \multicolumn{3}{p{16.635em}|}{46.96 (0.22)} \\
\hline
		
	\end{tabular}%
		\begin{tablenotes}[para,flushleft]
	\footnotesize
	\item $\beta^{*}$= proportion (\%) of outpatients who are assigned the other PHC; $\rho_i$ and $w_i$ indicate the average occupancy and the average wait time at the $i^{th}$ PHC subsystem, $i \in \{\text{doctor, pharmacy, NCD nurse, laboratory}\}$. 
    
\end{tablenotes}
\label{table2f}%
\end{threeparttable}
\end{table}%


The data in Table~\ref{table2f} also reveal consistent improvements across other operational outcomes following implementation of the RT-HFA algorithm. In particular, patient congestion at PHC$_2$ is significantly reduced, as evidenced by lower waiting times and resource occupancy levels across all queuing subsystems (NCD nurse, doctor, laboratory, and pharmacy). In summary, it is evident that assigning patients to healthcare facilities based on real-time LOS predictions can enhance operational efficiency and service quality across the healthcare network, benefiting both providers and patients. 



\noindent \underline{\textit{Behavioral considerations associated with the RT-HFA algorithm.}}

There is always a possibility that patients may not adhere to the RT-HFA algorithm. Note that results in Table~\ref{table2f} obtained from the implementation of the RT-HFA algorithm implicitly consider a 100\% compliance rate. To investigate how non-compliance will affect the overall operational outcomes, we conducted sensitivity analyses at varying compliance rates and generated simulation outcomes. An $x\%$ compliance rate was modeled by associating with each patient a Bernoulli random variable, where each patient complied with their facility recommendation with probability $x$ and did not comply with probability $1 - x$. We explored the impact of three scenarios involving adjusting compliance levels (i.e., setting the probability $x$) to: (a) 75\%, (b) 50\%, and (c) 25\%. 

\begin{figure}[htbp]
	\centering
	\begin{subfigure}{0.45\textwidth}
		\includegraphics[width=\linewidth]{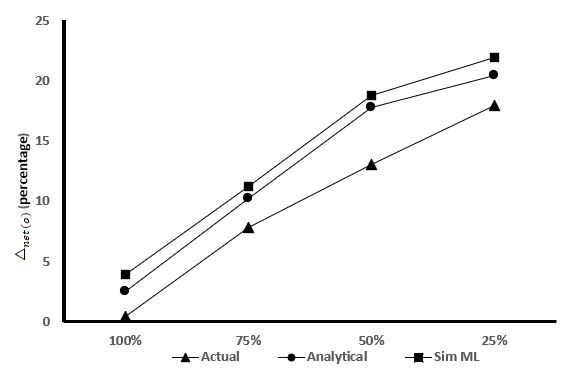}
		\caption{Doctor's occupancy.}
	\end{subfigure}
	\hspace{5pt}
	\begin{subfigure}{0.45\textwidth}
		\includegraphics[width=\linewidth]{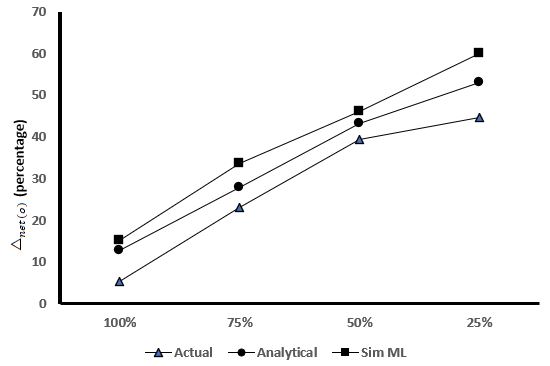}
		\caption{Wait time: doctor's subsystem.}
	\end{subfigure}
	\caption{$\Delta_{net(\rho-doc)}$ and $\Delta_{net(w)}$ at the doctor's queuing subsystem versus patient compliance with the facility recommendation generated by the RT-HFA algorithm.}
	\label{fig:sensitivity}
\end{figure}

\begin{figure}[htb]
	\centering
	\includegraphics[width=0.65 \textwidth]{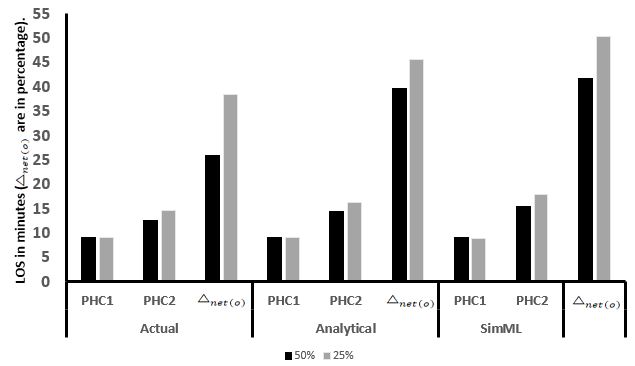}
	\caption{Variation in LOS estimates at PHC1 and PHC2 with 50\% and 25\% compliance levels with facility assignments.}
	\label{Fig7}
\end{figure}

We present outcomes from this sensitivity analysis for the outpatient queue in Figures~\ref{fig:sensitivity} and ~\ref{Fig7} where the y-axis denotes $\Delta_{net(o)}$ for $\rho_{doc}$ and $w_{opd}$, respectively. We observe that as the compliance level reduces, the resource occupancy levels, wait times, and average LOS values become less equitable across the network. For example, when all patients fully complied with the assignment algorithm, the $\Delta_{net(\rho-doc)}$ value was only 0.41\% (Table~\ref{table2f}). The $\Delta_{net(\rho-doc)}$ value increased significantly as compliance reduced: at 75\% compliance, the  increased to 7.82\%, and at 50\% compliance, it rose further to 13.02\%. A similar trend was recorded in wait times at the doctor's queuing subsystem and other PHC subsystems. 

Table~\ref{tab7} presents the variation in resource occupancies and wait times for all stations across the healthcare facilities at a 75\% compliance level. As compliance level decreases from 100\% to 75\%, the outcomes become more inequitable as evidenced by the $\Delta_{net(o)}$ values. Further, we observe that the expected LOS estimates at PHC$_2$ also start increasing: for example, when the assignment is done using actual LOS, it increases from 9.674 minutes to 10.358. 


\begin{table}[H]
  \centering
  \caption{Real-time healthcare facility assignment operational outcomes with a 75\% compliance rate.}
  \footnotesize
    \begin{tabular}{|l|l|p{6.215em}ll|}
    \hline
    \multicolumn{1}{|p{2.285em}|}{Cases} & \multicolumn{1}{p{5.785em}|}{Outcomes*} & \multicolumn{1}{p{6.215em}|}{PHC1} & \multicolumn{1}{p{6.5em}|}{PHC2} & \multicolumn{1}{p{7.215em}|}{$\Delta_{net(o)}$} \\
    \hline
    \multicolumn{1}{|l|}{{With RT-HFA}} & \multicolumn{1}{p{5.785em}|}{${\rho_{ncd}}$} & \multicolumn{1}{p{6.215em}|}{0.951 (0.007)} & \multicolumn{1}{p{6.5em}|}{1.001 (0.009)} & \multicolumn{1}{p{7.215em}|}{5.55\% (0.01)} \\
\cline{2-5}  (Actual LOS)        & \multicolumn{1}{p{5.785em}|}{${\rho_{phar}}$} & \multicolumn{1}{p{6.215em}|}{0.972 (0.004)} & \multicolumn{1}{p{6.5em}|}{1.116 (0.005)} & \multicolumn{1}{p{6.215em}|}{12.91\% (0.00)} \\
\cline{2-5}          & \multicolumn{1}{p{5.785em}|}{${\rho_{lab}}$} & \multicolumn{1}{p{6.215em}|}{0.754 (0.004)} & \multicolumn{1}{p{6.5em}|}{0.926 (0.004)} & \multicolumn{1}{p{6.215em}|}{18.55\% (0.00)} \\
\cline{2-5}          & \multicolumn{1}{p{5.785em}|}{${w_{phar}}$} & \multicolumn{1}{p{6.215em}|}{1.040 (0.034)} & \multicolumn{1}{p{6.5em}|}{1.628 (0.046)} & \multicolumn{1}{p{6.215em}|}{36.10\% (0.01)} \\
\cline{2-5}          & \multicolumn{1}{p{5.785em}|}{${w_{lab}}$} & \multicolumn{1}{p{6.215em}|}{1.093 (0.035)} & \multicolumn{1}{p{6.5em}|}{1.760 (0.046)} & \multicolumn{1}{p{6.215em}|}{37.06\% (0.01)} \\
\cline{2-5}          & \multicolumn{1}{p{5.785em}|}{${w_{ncd}}$} & \multicolumn{1}{p{6.215em}|}{1.009 (0.028)} & \multicolumn{1}{p{6.5em}|}{1.399 (0.032)} & \multicolumn{1}{p{6.215em}|}{27.82\% (0.02)} \\
\cline{2-5}          & \multicolumn{1}{p{5.785em}|}{${LOS}$} & \multicolumn{1}{p{6.215em}|}{9.218 (0.134)} & \multicolumn{1}{p{6.5em}|}{10.358 (0.103)} & \multicolumn{1}{p{6.215em}|}{11.00\% (0.01)} \\
\cline{2-5}          & $\beta^{*}$      & \multicolumn{3}{p{18.93em}|}{45.70 (0.23)} \\
    \hline
    \multicolumn{1}{|l|}{{With RT-HFA}} & \multicolumn{1}{p{5.785em}|}{${\rho_{ncd}}$} & \multicolumn{1}{p{6.215em}|}{0.947 (0.007)} & \multicolumn{1}{p{6.5em}|}{1.022 (0.015)} & \multicolumn{1}{p{6.215em}|}{7.33\% (0.01)} \\
\cline{2-5}    (AQT)       & \multicolumn{1}{p{5.785em}|}{${\rho_{phar}}$} & \multicolumn{1}{p{6.215em}|}{0.957 (0.005)} & \multicolumn{1}{p{6.5em}|}{1.129 (0.006)} & \multicolumn{1}{p{6.215em}|}{15.21\% (0.00)} \\
\cline{2-5}          & \multicolumn{1}{p{5.785em}|}{${w_{lab}}$} & \multicolumn{1}{p{6.215em}|}{0.735 (0.004)} & \multicolumn{1}{p{6.5em}|}{0.937 (0.006)} & \multicolumn{1}{p{6.215em}|}{21.56\% (0.01)} \\
\cline{2-5}          & \multicolumn{1}{p{5.785em}|}{${w_{phar}}$} & \multicolumn{1}{p{6.215em}|}{1.080 (0.036)} & \multicolumn{1}{p{6.5em}|}{1.718 (0.048)} & \multicolumn{1}{p{6.215em}|}{40.07\% (0.01)} \\
\cline{2-5}          & \multicolumn{1}{p{5.785em}|}{${w_{lab}}$} & \multicolumn{1}{p{6.215em}|}{0.990 (0.070)} & \multicolumn{1}{p{6.5em}|}{1.805 (0.057)} & \multicolumn{1}{p{6.215em}|}{43.89\% (0.02)} \\
\cline{2-5}          & \multicolumn{1}{p{5.785em}|}{${w_{ncd}}$} & \multicolumn{1}{p{6.215em}|}{0.974 (0.032)} & \multicolumn{1}{p{6.5em}|}{1.430 (0.034)} & \multicolumn{1}{p{6.215em}|}{31.86\% 0.02)} \\
\cline{2-5}          & \multicolumn{1}{p{5.785em}|}{${LOS}$} & \multicolumn{1}{p{6.215em}|}{9.102 (0.130)} & \multicolumn{1}{p{6.5em}|}{10.741 (0.145)} & \multicolumn{1}{p{6.215em}|}{15.26\% (0.01)                       } \\
\cline{2-5}          & $\beta^{*}$       & \multicolumn{3}{p{18.93em}|}{43.30 (0.24)} \\
    \hline
    \multicolumn{1}{|l|}{{With RT-HFA}} & \multicolumn{1}{p{5.785em}|}{${\rho_{ncd}}$} & \multicolumn{1}{p{6.215em}|}{0.933 (0.011)} & \multicolumn{1}{p{6.5em}|}{1.032 (0.010)} & \multicolumn{1}{p{6.215em}|}{9.58\% (0.01)} \\
\cline{2-5} (Sim-ML)         & \multicolumn{1}{p{5.785em}|}{${\rho_{phar}}$} & \multicolumn{1}{p{6.215em}|}{0.949 (0.006)} & \multicolumn{1}{p{6.5em}|}{1.139 (0.006)} & \multicolumn{1}{p{6.215em}|}{16.67\% (0.01)} \\
\cline{2-5}          & \multicolumn{1}{p{5.785em}|}{${w_{lab}}$} & \multicolumn{1}{p{6.215em}|}{0.723 (0.006)} & \multicolumn{1}{p{6.5em}|}{0.944 (0.006)} & \multicolumn{1}{p{6.215em}|}{23.37\% (0.01)} \\
\cline{2-5}          & \multicolumn{1}{p{5.785em}|}{${w_{phar}}$} & \multicolumn{1}{p{6.215em}|}{1.012 (0.038)} & \multicolumn{1}{p{6.5em}|}{1.737 (0.056)} & \multicolumn{1}{p{6.215em}|}{41.53\% (0.02)} \\
\cline{2-5}          & \multicolumn{1}{p{5.785em}|}{${w_{lab}}$} & \multicolumn{1}{p{6.215em}|}{0.986 (0.049)} & \multicolumn{1}{p{6.5em}|}{1.883 (0.063)} & \multicolumn{1}{p{6.215em}|}{46.18\% (0.02)} \\
\cline{2-5}          & \multicolumn{1}{p{5.785em}|}{${w_{ncd}}$} & \multicolumn{1}{p{6.215em}|}{0.933 (0.007)} & \multicolumn{1}{p{6.5em}|}{1.466 (0.028)} & \multicolumn{1}{p{6.215em}|}{32.26\% (0.01)} \\
\cline{2-5}          & \multicolumn{1}{p{5.785em}|}{${LOS}$} & \multicolumn{1}{p{6.215em}|}{9.089 (0.082)} & \multicolumn{1}{p{6.5em}|}{10.876 (0.154)} & \multicolumn{1}{p{6.215em}|}{16.43\% (0.01)} \\
\cline{2-5}          &  $\beta^{*}$     & \multicolumn{3}{p{18.93em}|}{42.97 (0.033)} \\
    \hline
    \end{tabular}%
    		\begin{tablenotes}[para,flushleft]
	\footnotesize
	\item $\beta^{*}$= proportion (\%) of outpatients who are assigned the other PHC; $\rho_i$ and $w_i$ indicate the average occupancy and the average wait time at the $i^{th}$ PHC subsystem, $i \in \{\text{doctor, pharmacy, NCD nurse, laboratory}\}$. 
    
\end{tablenotes}
  \label{tab7}%
\end{table}%

For the compliance levels of 50\% and 25\%, the operational outcomes exhibited a similar pattern, with decreasing equitability reflected by higher values of $\Delta_{net(o)}$. Figure \ref{Fig7} illustrates the variation in LOS estimates and the corresponding $\Delta_{net(o)}$ values across both PHCs. For PHC$_1$, LOS decreases marginally as the compliance level declines. In contrast, for PHC$_2$, a higher degree of noncompliance with the assignment algorithm results in fewer patients being routed to the less congested PHC$1$, leading to an increase in LOS. The $\Delta{net(o)}$ values reflect this behavior, indicating a widening disparity at both compliance levels. These results underscore the importance of adherence to assignment decisions and demonstrate how patient compliance with assignment decisions can significantly impact operational outcomes across the healthcare facility network.

\section{Discussion and Managerial Insights}
\label{rtlos_sec5}

Patient length of stay has been identified to be a key metric in diversion, assignment, and other healthcare facility performance improvement studies. This study presents a framework for the implementation of real-time healthcare facility assignment based on real-time length of stay predictions in a non-emergency setting, with assignment decisions made at the patient's point of origin. Such a framework is well-suited to healthcare networks in which patients are not \textit{a priori} assigned a first point of contact for medical care, and networks comprising facilities that serve a substantial number of walk-in patients. Such networks are, for example, ubiquitous in the Indian context \citep{rao2018quality}. The implementation of the framework is demonstrated using a discrete-event simulation of two primary healthcare facilities in the Indian context. Simulation results show that implementing the assignment algorithm balances patient loads across the network and leads to more equitable utilization of medical resources compared to scenarios without assignment. We also show empirically that adherence to the algorithm's recommendations is a prerequisite for effectively distributing patient loads across the network.


We develop an analytical queuing-theoretic method for estimating real-time LOS for use within the RT-HFA algorithm. The implementation of the framework with such predictors can be extended to facilities that can be modeled by a wide variety of queuing subsystems for which real-time delay prediction methods have been developed \citep{ibrahim2009real,ibrahim2018managing}. We also develop a Sim-ML methodology in which a validated DES model is used to generate system state data for training ML-based predictors. \citep{baldwa2020combined} also used a validated DES model to generate system state data for training ML-based classifiers for predicting whether or not patients will find a bed in a neurosurgical ward with a complex clinical condition based admissions algorithm within clinically relevant time thresholds. Our approach is well-suited for such queuing systems for which development of closed-form analytical expressions is intractable, and where system state information is either unavailable or not routinely recorded. In such settings, simulation-generated data provides a practical alternative for LOS prediction.

When comparing the AQT and Sim-ML methods in both prediction scenarios - predicting real-time LOS across the entire patient flow or at individual subsystems within the modeled facility - the AQT predictor either outperformed the Sim-ML predictor or demonstrated comparable performance. The AQT predictor that we develop is similar to a physics-based model of some system or phenomenon, in that it represents the most accurate description of the system that can be developed (assuming it is correct). Hence, the analytical predictor is likely to generate very accurate LOS estimates as long as the operating conditions of the real-world system deviate negligibly from those assumed in developing the analytical predictor. 

ML models typically do not directly utilize an understanding of the inner workings of the systems but use data regarding key system variables and the prediction variable to identify operational patterns related to LOS values. Hence, it is not surprising that the analytical predictors perform better than certain ML models. That said, certain ML models - notably artificial neural networks and extreme gradient-boosted trees - also outperform the analytical predictor. This is not surprising, given the universal approximation properties of neural networks that facilitate their capture of complex nonlinear system behavior \citep{bishop2006pattern}. Ensemble decision tree ML models also possess similar capabilities and have been known to perform comparably or better than ANNs with tabular data (as is the case here) \citep{shwartz2022tabular}. Therefore, the question of whether the Sim-ML approach or the analytical approach works better depends on the ML model in question and the extent of uncertainty in the operating conditions of the real-world system in question. 

The ease of deployment in terms of the number of system state variables needed to be monitored is likely to be similar for both RT-LOS prediction methods. However, retraining and deploying an ML method on new data if operating conditions change is likely to be considerably easier than developing a new AQT method, especially if the change in system operating conditions is substantial. This is because while a validated DES of the system may be required to initiate the development of the ML RT-LOS predictor, its deployment will regardless prospective recording of adequate system state data. Therefore, as the deployment of the ML predictor matures, a comprehensive dataset (system state and LOS information) is likely to be assembled, and the use of the DES is unlikely to be required for future retraining of the ML RT-LOS predictor. However, DESs have uses beyond facilitating RT-LOS prediction, and maintaining and using a system DES is likely to continue to yield actionable operational insights \citep{law2007simulation}.

An important question is how the assignment algorithm from Algorithm~\ref{alg:the_alg} would scale in computational and operational terms as the healthcare facility network size grows. Before deploying this approach in larger networks, it is essential to identify which facilities should generate real-time LOS predictions for assignment decisions. For instance, in a network with 1,000 facilities, it may not be necessary to generate predictions at every location, especially those far from the patient. Defining the minimal set of facilities for assignment consideration is a key area for future research to enable scalability in large networks.


\noindent \underline{\textit{Managerial Insights}}

\begin{itemize}
    \item Our approach can be utilized in two ways by healthcare administrations: (a) to inform patients about their expected LOS estimates at different facilities in the network, especially in the Indian context, and (b) to aid patients in their decision-making regarding which healthcare facility to visit, where care can be received in the smallest time duration.
    \item Healthcare facility networks planning to implement an assignment algorithm can consider assignment based on real-time LOS prediction, especially in a non-emergency setting, given that it is likely to result in more equitable utilization of resources and balancing patient load across the network.
    \item The deployment of such a facility assignment method across a network requires an information technology (IT) system to which each facility in the network is linked, and shares the system state data required for the RT-HFA algorithm. Such IT systems are present in public health systems in many countries: for example, the NHSQuicker system in the United Kingdom informs patients regarding average waiting times in emergency care facilities in a public health network \citep{mustafee2020providing}.
\end{itemize}

\bibliographystyle{model5-names}\biboptions{authoryear}
\bibliography{cas-refs}





\end{document}